\documentclass[final]{siamltex}

\usepackage{epsf}
\usepackage{amssymb}
\usepackage{amsbsy}
\usepackage{verbatim}
\usepackage{amsmath}
\usepackage{tabularx}

\usepackage{graphicx}

\usepackage{color}

\usepackage{datetime}

\usepackage[colorlinks=true, linkcolor=blue, citecolor=blue, filecolor=blue, runcolor=blue, urlcolor = blue]{hyperref}

\newcommand{\OpFldDissp}{\mathcal{L}}
\newcommand{\OpFSDissp}{\Upsilon}
\newcommand{\OpFSCouple}{\Gamma}
\newcommand{\OpSFCouple}{\Lambda}

\newcommand{\RegAdjoint}{{T}}

\newcommand{\proj}{\wp}

\newcommand{\Eng}{\Phi}

\newcommand{\mb}[1]{\mathbf{#1}}

\newcommand{\subtxt}[1]{ {\mbox{\tiny #1}} }

\newcommand{\bkFunc}{u}
\newcommand{\avgFunc}{f}

\newcommand{\p}{\partial}
\newcommand{\f}{\frac}

\newcommand{\vOrth}{\beta}

\ifpdf
\fi

\begin{document}

\title{Stochastic Reductions for Inertial Fluid-Structure Interactions subject to Thermal Fluctuations}

\author{Gil Tabak 
\thanks{Department of Applied Physics, Stanford University.
E-mail: tabakg@stanford.edu;
Work supported by NSF CAREER DMS - 0956210, undergraduate participant in 
NSF REU DMS-0852065 and the CCS program at UCSB.
Graduate support from
 the Department of Defense (DoD) through the National Defense Science \& Engineering Graduate Fellowship (NDSEG) Program 
and the Stanford Graduate Fellowship.
}  
\and Paul J. Atzberger
\thanks{Department of Mathematics, University of California Santa Barbara.
Mail: 6712 South Hall, Santa Barbara, CA 93106; 
E-mail: atzberg@math.ucsb.edu; phone: 805-893-3239;
Work supported by NSF CAREER Grant DMS - 0956210 and DOE CM4. URL:
\href{http://atzberger.org/}{
$\mbox{http://atzberger.org/}$}
}
}

\maketitle

\begin{abstract}
We investigate the dynamics of elastic microstructures within a fluid that are subjected to thermal fluctuations.  
We perform 
analysis to obtain systematically simplified descriptions of the mechanics in the limiting regimes when 
(i) the coupling forces that transfer momentum between the fluid and microstructures is strong, 
(ii) the mass of the microstructures is small relative to the displaced mass of the fluid, 
and (iii) the response to stresses results in hydrodynamics that relax rapidly to 
a quasi-steady-state relative to the motions of the microstructure.
We derive effective equations using a singular perturbation analysis of the Backward 
Kolmogorov equations of the stochastic process.
Our continuum mechanics description is based on the Stochastic Eulerian Lagrangian Method (SELM)
which provides a framework for approximation of the fluid-structure interactions when subject 
to thermal fluctuations.  
\end{abstract}


\pagestyle{myheadings}
\thispagestyle{plain}
\markboth{G. TABAK AND P. J. ATZBERGER}
{STOCHASTIC REDUCTION OF SELM}

\section{Introduction}

In many applications an important role is played by the mechanics of spatially extended elastic bodies within a fluid that are subjected to thermal fluctuations.  At a microscopic level, elastic bodies may represent the polymers of a complex fluid~\cite{Doi1986,Bird1987Vol2}, the individual or collective behaviors of amphiphilic molecules of a lipid bilayer membrane~\cite{Helfrich1973,Lipowsky1991,Cai1995}, the cilia or flagellum driving the swimming of microorganisms~\cite{Lauga2009,Purcell1977}, or the components of micro-mechanical devices~\cite{Raman2008,Ho1998}.  Even in the absence of thermal fluctuations, the mechanics of fluid-structure interactions pose a number of difficult and long-standing challenges.  The mechanics can exhibit rich behaviors through the interplay of the fluid flow and elastic stresses.  The fluid flow can drive deformations of the elastic structures while at the same time the elastic stresses of the deformed bodies can feedback to resist flow of the fluid or even drive flow.  
This can result in 
complicated
long-range coupling between the motions of different bodies or the 
motions between different parts of a single spatially extended body.  When subject to thermal fluctuations, this manifests as diffusive motions with long-range correlations between the immersed microstructures.   In continuum mechanics it is natural to consider descriptions for these interactions in which explicit boundary conditions are formulated at the interface between the elastic body with the fluid (i.e. the traction stress and kinematic conditions).  However, in practice this approach is often intractable for physical analysis and prohibitively expensive in computational simulations.  

To obtain more tractable descriptions, while still capturing essential features of the mechanics, many approximations of the fluid-structure interactions have been developed.  These include the Immersed Boundary Method~\cite{Peskin2002,Atzberger2007a,Kim2007a, Donev2012}, Arbitrary Lagrangian-Eulerian Methods~\cite{Braescu2007, Duarte2004}, Stokesian-Brownian Dynamics~\cite{Brady1988}, Lattice-Boltzmann Method~\cite{Duenweg2007, Duenweg2008}, and Force Coupling Method~\cite{Maxey2001,Climent2003}.  While these approximations may perform adequately in the deterministic setting, when introducing the thermal fluctuations the approximations can often introduce unphysical dissipation, spurious drifts, and other artifacts affecting the propagation of fluctuations throughout the mechanical system.  
The Stochastic Eulerian Lagrangian Method (SELM) has been introduced in~\cite{AtzbergerSELM2011} to provide a framework for the development of approximate fluid-structure coupling schemes that are amenable to thermal fluctuations.  Provided a few criteria are satisfied by the approximate fluid-structure coupling, SELM provides an approach to formulate equations of motion for the fluid-structure system and to introduce stochastic driving fields that yield equilibrium fluctuations having Gibbs-Boltzmann statistics and stochastic dynamics satisfying detailed balance~\cite{AtzbergerSELM2011}.  

The SELM equations are formulated at the level of inertial fluid-structure interactions and allow for the case of slip between the microstructure and fluid.  However, many of the widely used approximation methods for fluid-structure interactions are formulated in other physical regimes of interest, such as in the limit with strong fluid-structure coupling with no-slip, small body mass, or rapid hydrodynamic relaxation~\cite{AtzbergerSELM2011}. 
A primary objective of our present work is to develop reductions of the SELM equations that are directly applicable to these regimes and to compare the reduced descriptions with well-known results in fluid mechanics and statistical mechanics.  Some of this work was carried out partially based on formal calculations in our prior paper~\cite{AtzbergerSELM2011}.  The focus of the current work is (i) to develop a more systematic and reliable approach to reduce the SELM equations, (ii) to characterize precisely each of the physical regimes using 
dimensional analysis, and (iii) to consider an important previously unexplored limit in which the fluid-structure coupling forces becomes strong to yield no-slip.

Our approach is based on a singular perturbation analysis of the Backward Kolmogorov equation (BKE) carried out using formal methods similar to those 
in~\cite{Kramer2004,Majda2004a,Majda2003, Papanicolaou1976,Kurtz1973}.  We present a formulation of the fluid-structure mechanics based on the Stochastic Eulerian Lagrangian Method (SELM) in Section~\ref{sec_SELM_method}.  We present dimensional analysis of the SELM equations in Section~\ref{sec_dimension_analysis}.  We summarize our main results for each physical regime in Section~\ref{sec_reduction}.  The general singular perturbation is developed in Section~\ref{sec_general_reduction_procedure}.  We present the detailed derivations of the reduced equations in Section~\ref{sec_derivation_reduction}.  The reduced equations provide connections between different physical regimes for many common approximate models of fluid-structure interactions when subject to thermal fluctuations.  The results could also be useful to eliminate sources of numerical stiffness, providing ways to develop efficient methods for computational simulations.

\section{Stochastic Eulerian Lagrangian Method : Inertial Regime}
\label{sec_SELM_method}
The fluid-structure interactions are described by the fluctuating hydrodynamic equations
\begin{eqnarray}
\label{fld_struct_u_v_x}
\rho \frac{d \mb{u}}{dt} & = & \mu \Delta \mb{u} -\nabla{p} + \OpSFCouple[\OpFSDissp(\mb{v}-\OpFSCouple \mb{u})] + \mb{f}_{\text{thm}} \\
\nabla\cdot\mb{u}        & = & 0.
\end{eqnarray}
They are coupled to the elastic structure equations
\begin{eqnarray}
\label{fld_struct_x}
m \frac{d\mb{v}}{dt}     & = & -\OpFSDissp(\mb{v} - \OpFSCouple \mb{u})- \nabla_\mb{X} \Phi[\mb{X}] + \zeta +  \mb{F}_{\text{thm}} \\
\frac{d\mb{X}}{dt}       & = & \mb{v}.
\end{eqnarray}
The thermal fluctuations are accounted for by Gaussian stochastic fields $\mb{f}_{\text{thm}}$ and $\mb{F}_{\text{thm}}$ 
with mean zero and covariances
\begin{eqnarray}
\label{equ_SELM_thermal1_0}
\langle \mathbf{f}_\subtxt{thm}(s)\mathbf{f}_\subtxt{thm}^{\RegAdjoint}(t) \rangle & = & -\left(2k_B{T}\right)\left(\mu\Delta - \OpSFCouple\OpFSDissp\OpFSCouple\right)\hspace{0.03cm}\delta(t - s) \\
\label{equ_SELM_thermal2_0}
\langle \mathbf{F}_\subtxt{thm}(s)\mathbf{F}_\subtxt{thm}^{\RegAdjoint}(t) \rangle & = & \left(2k_B{T}\right)\OpFSDissp\hspace{0.03cm}\delta(t - s)\\
\label{equ_SELM_thermal3_0}
\langle \mathbf{f}_\subtxt{thm}(s)\mathbf{F}_\subtxt{thm}^{\RegAdjoint}(t) \rangle & = & -\left(2k_B{T}\right)\OpSFCouple\OpFSDissp\hspace{0.03cm}\delta(t - s).
\end{eqnarray}
In the notation, the $\mb{a}\mb{b}^T$ denotes the tensor product of vector fields $\mb{a}$ and $\mb{b}$, 
$k_B{T}$ the thermal energy, $\delta(t - s)$ the Dirac delta function, and $\langle \cdot \rangle$ 
a probability expectation of the random fields.
The fluid velocity is a field given by $\mb{u}$, and the structure configuration and velocities are  $N$-dimensional vectors given
by $\mb{X}(\mb{q})$ and $\mb{v}(\mb{q})$ with $\mb{q}$ the parameterization of the structure.  
The $\Phi$ denotes the potential energy associated with the structure configuration.   The $\rho$ denotes the fluid density, $p$ the pressure, $\mu$ the dynamic fluid viscosity, 
and $m$ the excess mass of the structure. $\zeta$ are some constraint forces on the microstructures which we assume are holonomic.

We remark that the full material derivative $d\mathbf{u}/dt = \partial \mathbf{u} /\partial t + \mathbf{u}\cdot\nabla \mathbf{u}$ is not well-defined in the stochastic setting since the 
fluctuating fluid velocity $\mathbf{u}$ is highly irregular and lacks a definition in the point-wise sense
requiring an interpretation using generalized functions (distributions).  As a consequence, the product $\mathbf{u}\cdot\nabla \mathbf{u}$ representing the advection of momentum is not well defined for such $\mathbf{u}$.  To avoid these issues, in our model of the fluid dynamics we treat in the present work $d\mathbf{u}/dt = \partial \mathbf{u} /\partial t$.  For a more in-depth discussion, see~\cite{AtzbergerSELM2011}. 

We remark that in equation~\ref{fld_struct_x} we assume the elastic body is described by generalized coordinates $\mathbf{X},\mathbf{v}$ involving either a finite or countably infinite number of degrees of freedom.  We emphasize that equation~\ref{fld_struct_x} is not a continuum elasticity equation, although such a generalization can be straight-forwardly formulated for SELM.  While this equation is most natural for modeling a collection of interacting point particles, the equations also can be motivated as a starting point for numerical simulations for continuum elastic bodies when one considers a semi-discretized representation such as finite elements. 
  
The fluid-structure interactions and associated momentum exchange is modeled by the terms 
$-\OpFSDissp(\mb{v} - \OpFSCouple \mb{u})$ and $\OpSFCouple[\OpFSDissp(\mb{v}-\OpFSCouple \mb{u})]$.  
The first term accounts for drag force exerted on a structure by the fluid.  The second term
represents the spatial distribution of the equal-and-opposite forces (stresses) that 
act on the fluid.  The operators $\OpSFCouple, \OpFSCouple$ transfer information between the 
Eulerian and Lagrangian descriptions.  The $\OpFSCouple$ determines from the state of the fluid
a local reference velocity to which to compare in determining the local drag exerted on the structures.   The $\OpSFCouple$
accounts for how the equal-and-opposite drag force exerted by the structures are distributed spatially within the fluid body.  We assume that the linear operators $\OpFSCouple$ and $\OpSFCouple$ may depend on the microstructural configuration $\mathbf{X}$, but $\OpFSDissp$ is constant in time.  
We remark that for a finite collection of $N$ particles in $d$-dimensional space the operator distributing force to the fluid body can be thought of as $\Lambda = \Lambda[\mb{X}] : \mathbb{R}^{Nd} \rightarrow (C(\mathbb{R}^d))^d$, where $C(\Omega)$ is the space of real-valued continuous functions on $\Omega$ and the superscript $(\cdot)^d$ indicates the product space so that $f\in (C(\Omega))^k$ means that $f : \Omega \rightarrow \mathbb{R}^k$.  Similarly, $\Gamma[\mb{X}] : (C(\mathbb{R}^d))^d \rightarrow \mathbb{R}^{Nd}$.
A number of natural conditions arise on these operators to ensure that dissipation in the coupling occurs only through the 
$\OpFSDissp$-drag and not as a consequence of interconversion between the Eulerian and Lagrangian reference frames.  
This requires the condition that the operators satisfy the adjoint condition $\OpSFCouple = \OpFSCouple^T$ in the sense that
\begin{eqnarray*}
\int(\OpFSCouple \mathbf{u}) (\mathbf{q}) \cdot \mathbf{v}(\mathbf{q}) d\mathbf{q} 
= \int \mathbf{u}(\mathbf{x}) \cdot (\OpSFCouple \mathbf{v})(\mathbf{x}) d\mathbf{x}
\end{eqnarray*}
for any fluid-field $\mathbf{u}$ and structure velocity $\mathbf{v}$ (see~\cite{Peskin2002,AtzbergerSELM2011}).  The $\mb{q}$ denotes the general ''index'' used to parametrize the microstructure configuration.
Another natural condition which ensures the drag is dissipative is that the 
operator $\OpFSDissp$ is positive semi-definite.  
The general formulation given in~\ref{fld_struct_u_v_x} -- \ref{equ_SELM_thermal3_0} 
is referred to throughout as the Stochastic Eulerian Lagrangian Method.  For more details concerning the motivation of these equations 
and their derivation, see~\cite{AtzbergerSELM2011}.  

In practice, many different choices can be made for the coupling operators $\OpSFCouple$, $\OpFSCouple$, 
and $\OpFSDissp$.  For the purposes of our analysis, we leave this choice general.  However,
for concreteness we mention that a common approach is to use local Stokesian drag and the fluid-structure 
coupling scheme of the \textit{Immersed Boundary Method} given in~\cite{Peskin2002}.
This would give the specific coupling operators
\begin{eqnarray}
\nonumber
\left(\OpFSCouple \mb{u}\right)(\mb{q})  =  \int \delta_a(\mb{y} - \mb{X}(\mb{q})) \mb{u}(\mb{y},t) d\mb{y}, \mbox{\hspace{0.3cm}}
\left(\OpSFCouple \mb{F}(\mb{q})\right)(\mb{y})  =  \delta_a(\mb{y} - \mb{X}(\mb{q})) \mb{F}(\mb{q}), \mbox{\hspace{0.3cm}} 
\OpFSDissp                               =  6\pi\mu R.
\end{eqnarray}

The $\delta_a$ is a special kernel function localized around the structure designed to have desirable
numerical properties that preserve to a good approximation translation invariance of the structure dynamics
despite the breaking of this symmetry by the numerical discretization lattice of the fluid.   We refer to the specific choice given in~\cite{Peskin2002} as the Peskin $\delta$-function.  In the notation, the $R$ is the 
hydrodynamic radius attributed to the effective local size of the structure.    
More general coupling schemes can also be developed to which our presented analysis is applicable.  
For more details on how these fluid-structure approaches are used in practice, 
see~\cite{AtzbergerSELM2011,Atzberger2007a,Peskin2002}.

\subsection{Reformulation in Terms of the Total Momentum Density Field}
\label{sec_reform_total_mom}
For our analysis it is useful to reformulate the equations in terms of a 
total momentum density field.  This field accounts simultaneously for both the momentum of the fluid 
and structures at a given location in space.  For this purpose we define the \textit{total momentum density field}
\begin{eqnarray}
\label{def_total_mom}
\mb{p}(\mb{x},t) = \rho \mb{u}(\mb{x},t) + \OpSFCouple\left\lbrack m\mb{v}(t) \right\rbrack.   
\end{eqnarray}
Also for convenience, 
and to treat more general approaches for the fluid-structure system, we introduce the operator
$\OpFldDissp = \mu \Delta$ in place of the Newtonian stress term.  We only assume $\OpFldDissp$ is 
negative semi-definite throughout.   This allows for fluid-structure interactions to be expressed 
in terms of $\mb{p}, \mb{v}, \mb{X}$ as 
\begin{eqnarray}
\label{equ_SELM_I_1}
\frac{d\mathbf{p}}{dt}  & = & \rho^{-1} \OpFldDissp \left(\mathbf{p} - \OpSFCouple[m\mathbf{v}]\right) + \OpSFCouple[-\nabla_{\mathbf{X}}\Phi(\mathbf{X})] 
                         + \left(\nabla_{\mathbf{X}} \OpSFCouple[m\mathbf{v}]\right)\cdot \mathbf{v} 
                         + \lambda + \mathbf{g}_\subtxt{thm} \\
\label{equ_SELM_I_2}
m\frac{d\mathbf{v}}{dt} & = & -\OpFSDissp \mathbf{v} 
						+ \rho^{-1} \OpFSDissp \OpFSCouple {\left(\mathbf{p} - \OpSFCouple[m \mathbf{v}]\right)} 
                            -\nabla_{\mathbf{X}}\Phi(\mathbf{X})
                            + \zeta + \mathbf{F}_\subtxt{thm} \\
\label{equ_SELM_I_3}
\frac{d\mathbf{X}}{dt}  & = & \mathbf{v}.
\end{eqnarray}
The thermal fluctuations are given by the Gaussian stochastic fields with covariance
\begin{eqnarray}
\langle \mathbf{g}_\subtxt{thm}(s)\mathbf{g}_\subtxt{thm}^{\RegAdjoint}(t) \rangle = -\left(2k_B{T}\right)\OpFldDissp\hspace{0.06cm}\delta(t - s), \mbox{\hspace{0.25cm}}
\langle \mathbf{F}_\subtxt{thm}(s)\mathbf{F}_\subtxt{thm}^{\RegAdjoint}(t) \rangle =  \left(2k_B{T}\right)\OpFSDissp\hspace{0.06cm}\delta(t - s) 
\mbox{\hspace{0.25cm}}\\
\langle \mathbf{g}_\subtxt{thm}(s)\mathbf{F}_\subtxt{thm}^{\RegAdjoint}(t) \rangle =  0. \mbox{\hspace{8.4cm}}
\label{equ_SELM_I_6}
\end{eqnarray}
The $\lambda$ and $\zeta$ are Lagrange multipliers enforcing holonomic constraints on the system 
(such as incompressibility of the fluid or rigid-body restrictions on deformations of the structures).
The stochastic driving field for the total momentum is related to the stochastic driving
fields of the fluid and structures in equations~\ref{fld_struct_u_v_x} and~\ref{fld_struct_x}
by $\mathbf{g}_\subtxt{thm} = \mathbf{f}_\subtxt{thm} + \OpSFCouple[\mathbf{F}_\subtxt{thm}]$, see \cite{AtzbergerSELM2011}.

An important technical issue is that the total momentum field $\mathbf{p}$ itself need not be incompressible (solenoidal) as
a consequence of the somewhat arbitrary way momentum of the structures is spatial distributed by 
$\OpSFCouple\left\lbrack m\mb{v}(t) \right\rbrack$.  This will require some care in our calculations to ensure that the 
incompressibility constraint is satisfied by the corresponding fluid velocity field $\mathbf{u}$.  We handle this technical
point by considering a decomposition of the total momentum field into a solenoidal part and non-solenoidal part.  An important
feature is that only the solenoidal part of the total momentum field plays a significant role in the coupled 
fluid-structure equations~\ref{equ_SELM_I_1} -- \ref{equ_SELM_I_6}. 

\subsection{Handling Constraints : Incompressibility}
\label{sec_incorporating_incompressibility}
An important feature of the microstructure equations is that they depend on the total momentum field $\mb{p}$ only 
through the velocity field $\mathbf{u} = \mathbf{p} - \OpSFCouple[m\mathbf{v}]$ which is constrained to be solenoidal 
(incompressible).  This can be expressed by the requirement that $\mathbf{u} = \proj \mathbf{u} = \proj\mathbf{p} - \proj\OpSFCouple[m\mathbf{v}]$, 
where $\proj$ is the projection operator of a field to its solenoidal part.  This well-known approach for expressing the incompressibility of 
hydrodynamic equations~\cite{Chorin1968} yields for SELM the following closed set of equations
\begin{eqnarray}
\label{equ_SELM_inc_I_1}
\frac{d(\proj \mathbf{p}) }{dt}  & = & \proj \bigg[ \rho^{-1} \OpFldDissp \left(\mathbf{p} - \OpSFCouple[m\mathbf{v}]\right) + \OpSFCouple[-\nabla_{\mathbf{X}}\Phi(\mathbf{X})] 
                         + \left(\nabla_{\mathbf{X}} \OpSFCouple[m\mathbf{v}]\right)\cdot \mathbf{v} 
                          + \mathbf{g}_\subtxt{thm} \bigg]\\
\label{equ_SELM_inc_I_2}
m\frac{d\mathbf{v}}{dt} & = & -\OpFSDissp \mathbf{v} 
						+ \rho^{-1} \OpFSDissp \OpFSCouple \proj {\left(\mathbf{p} - \OpSFCouple[m \mathbf{v}]\right)} 
                            -\nabla_{\mathbf{X}}\Phi(\mathbf{X})
                             + \mathbf{F}_\subtxt{thm} \\
\label{equ_SELM_inc_I_3}
\frac{d\mathbf{X}}{dt}  & = & \mathbf{v}.
\end{eqnarray}
This means that only the dynamics of $\proj \mathbf{p}$ need be considered explicitly since the non-solenoidal components play no role in the fluid-structure dynamics.  The equations~\ref{equ_SELM_inc_I_1}--\ref{equ_SELM_inc_I_3} and the covariance of $\mathbf{g}_{\text{thm}}$ take the same identical form as equations~\ref{equ_SELM_I_1}--\ref{equ_SELM_I_6} when making the substitutions
\begin{align}
\tilde{\mathbf{p}} = \proj \mathbf{p}, &&
\tilde{\OpFldDissp} = \proj \OpFldDissp, &&
\tilde{\OpSFCouple} = \proj \OpSFCouple, &&
\tilde{\OpFSCouple} =  \OpFSCouple \proj. &&
\end{align}
We use a similar approach to simplify the handing of the holonomic constraints  for the 
microstructures by using appropriate generalized coordinates for which the constraints do not occur, and hence drop the $\zeta$ term.
To avoid clutter in our notation when carrying out our reductions, we shall use these conventions and not explicitly distinguish the 
case with such constraints.

\section{Dimensional Analysis} 
\label{sec_dimension_analysis}
The SELM fluid-structure dynamics exhibit a broad range of spatial/temporal scales.  To characterize these scales 
and identify interesting limiting physical regimes, we perform a dimensional analysis of the SELM 
equations~\ref{equ_SELM_I_1} -- \ref{equ_SELM_I_6}.  The Buckingham $\Pi$-theorem states that for a physical system 
with $M$ parameters and $m$ fundamental physical units the system state can be characterized by $M - m$ non-dimensional 
groups $\Pi_1, \cdots, \Pi_{M-m}$~\cite{Buckingham1914, Barenblatt1996}.  
 
For the SELM equations there are $M = 7$ independent parameters $\ell, m, \rho, \mu, k_B T, \Upsilon_0$, and $\Phi_0$.  The $\ell$ is the effective size of particles/micro-structures used in the coupling operators, $m$ is the excess mass of the microstructures, $\rho$ is the fluid density, $\mu$ is the fluid dynamic viscosity, $k_B{T}$ is the thermal energy, $\Upsilon_0$ is the microstructure-fluid drag coefficient, and $\Phi_0$ is the interaction energy between the microstructures. The other symbols that appear in SELM depend on these parameters.  Throughout our discussion of the operators that appear in equation~\ref{equ_def_consts2} and for the purposes of our dimensional analysis, we express an operator $\mathcal{O}$ in terms of a dimensional scalar parameter characterizing  the magnitude of the operator $\mathcal{O}_0$ and a non-dimensional operator $\bar{\mathcal{O}}$ so that
$\mathcal{O} = \mathcal{O}_0 \bar{\mathcal{O}}$.
For instance, the linear operator $\OpSFCouple = \OpSFCouple_0 \bar{\OpSFCouple} = ({1}/{\ell^3}) \bar{\OpSFCouple} \left[ {\mathbf{X}}/{\ell}\right], $ where the key parameter $\Lambda_0 = {1}/{\ell^3}$ characterizes the magnitude and the linear non-dimensional operator is $\bar{\Lambda}\left[\bar{\mb{X}}\right] = \Lambda_0^{-1}\Lambda\left[\ell\bar{\mb{X}}\right]$.  We remark that the viscosity parameter $\mu$ plays a similar role, for instance, when the operator is $\OpFldDissp = \mu \Delta = ({\mu }/{\ell^2}) \bar{\Delta} = \mathcal{L}_0 \bar{\mathcal{L}}$ where $\mathcal{L}_0 = ({\mu }/{\ell^2})$.  The other operators are treated similarly.  We make an additional remark concerning the dimensional analysis and our treatment of the temperature of the system $k_B{T}$.  We note that throughout SELM, the temperature only contributes through the term $k_B T$ which has units of energy setting the scale of thermal contributions which warrants that the term be treated as one of the parameters as opposed to an independent physical unit~\cite{Barenblatt1996}.  From the dynamical symmetry of physical laws, our parameters, and the unit system, there are for this formulation of SELM only four non-dimensional groups $\Pi_1, \Pi_2, \Pi_3, \Pi_4$~\cite{Buckingham1914, Barenblatt1996}.  To obtain a useful set of non-dimensional groups, we introduce specific characteristic scales for the system in Section~\ref{sec_Char_Scales}.

\subsection{Characteristic Scales}
\label{sec_Char_Scales}
An important time-scale on which the particle velocity relaxes is $\tau_v = m/\OpFSDissp_0$.  
The $\OpFSDissp_0$ can be interpreted as the characteristic strength of the momentum-exchange coupling and 
$m$ is the excess mass of the microstructure.  This time-scale is similar to the one considered when 
reducing the Langevin equations to the Smoluchowski equations~\cite{Reichl1998, Gardiner1985}.  
Another important time-scale is $\tau_k = \sqrt{m_0 \ell^2/k_B{T}}$ which gives 
the time duration for a fluid parcel to move the distance $\ell$ when it has kinetic 
energy $k_B{T}$.  We remark that the parameters $\tau_v$ and $\tau_k$ are formulated for convenience and
combinations of the previously listed parameters so are already characterized by the dimensional analysis of the previous Section~\ref{sec_dimension_analysis}. The fundamental length scale $\ell$ characterizes the spatial scale of a fluid parcel directly influenced by a particle as represented by the coupling operator $\Lambda = \Gamma^T$.  
For instance, the Immersed Boundary Method coupling~\cite{Peskin2002,Atzberger2007a} for a particle has the form $\Gamma{\mb{u}} = \int \delta_{a}(\mb{x} - \mb{X}) \mb{u}(\mb{x},t) d\mb{x}$ and $\Lambda{\mb{F}} = \delta_{a}(\mb{x} - \mb{X}) \mb{F}$.  The Peskin $\delta_a$-function has finite support contained within a sphere of radius $3a$ around the origin~\cite{Peskin2002,Atzberger2007a}.  We identify the parcel length-scale as $\ell \sim a$.  We define the characteristic mass of the fluid parcel $m_0 = \rho \ell^3$.
This defines the two time-scales and provides a natural way to characterize the strength of the fluid-structure coupling
through the non-dimensional group 
\begin{eqnarray}
\label{equ_def_epsilon}
\epsilon = \tau_v/\tau_k, \mbox{\hspace{0.25cm} with $\Pi_1 = \epsilon$}.
\end{eqnarray}
This ratio characterizes how long it takes the microstructure momentum 
to decorrelate relative to the microstructure moving a significant distance.
The momentum-exchange coupling is strong when $\epsilon$ is small.

To characterize inertial effects, we consider the magnitude of the microstructure excess mass $m$
and the mass of fluid displaced by the microstructure to obtain the non-dimensional group 
\begin{eqnarray}
\label{equ_def_kappa_inv}
\kappa^{-1} = m/\rho\ell^3 = m / m_0, \mbox{\hspace{0.25cm} with $\Pi_2 = \kappa^{-1}$}.
\end{eqnarray}
The reference displaced mass is $m_0 = \rho\ell^3$.  Inertial effects of the fluid body dominate those of the microstructure when $\kappa^{-1}$ is small.  

The mechanics of the microstructure depend importantly on the potential energy $\Phi = \Phi_0 \bar{\Phi}\left( {\mathbf{X}}/{\ell} \right)$ of a configuration, where $\Phi_0$ is its characteristic scale and $\bar{\Phi}$ is non-dimensional. 
We characterize the importance of the microstructure potential energy relative to thermal energy using the non-dimensional group
\begin{align}
\alpha = \Phi_0/k_B T, \mbox{\hspace{0.25cm} with $\Pi_3 = \alpha$}.
\end{align}
We shall impose throughout our analysis the condition that $\alpha \kappa = O(1)$ when performing asymptotic analysis and for concreteness make the simplification throughout that $\alpha \kappa = 1$.  This condition ensures that the conservative forces contribute comparably to other influences on the dynamics of the microstructure.  

I
It is useful to define the following non-dimensional group that plays a role similar to the Reynolds number
\begin{eqnarray}
\label{equ_def_delta}
\delta = \rho \ell U/\mu, 
\mbox{\hspace{0.25cm} with $\Pi_4  = \delta$}.
\end{eqnarray}
The $U$ is a characteristic velocity scale of the fluid. Throughout  $U = \ell / \tau_k$. 
This term is used mainly in our analysis for the limit of rapid hydrodynamic relaxation developed in  Section~\ref{sec_derivation_rapid_fluid_relax}.

Using these parameter groups and characteristic scales we can express the SELM equations in non-dimensional form.  From the Buckingham-$\Pi$ Theorem, we know that all parameters of the physical system can be expressed
in terms of these four non-dimensional groups $\epsilon,\kappa^{-1},\alpha,\delta$~\cite{Buckingham1914, Barenblatt1996}.
To obtain readily 
such expressions,
we nondimensionalize the variables
\begin{align}
\label{equ_def_consts1}
t = \tau_{k} \bar{t}, 
 \hspace{0.595in}
\mathbf{X} = \ell \mathbf{\bar{X}}, 
 \hspace{0.595in}
\mathbf{v} = v_0 \mathbf{\bar{v}} = \frac{\ell}{\tau_k}  \mathbf{\bar{v}},  
 \hspace{0.595in}
\mathbf{p} = p_0 \mathbf{\bar{p}} = \frac{m_0}{\tau_k \ell^2} \mathbf{\bar{p}},
\\
\label{equ_def_consts2}
\OpSFCouple = \OpSFCouple_0 \bar{\OpSFCouple} \left( \frac{\mathbf{X}}{\ell}\right)= \frac{1}{\ell^3} \bar{\OpSFCouple} \left( \frac{\mathbf{X}}{\ell}\right),  
 \hspace{0.2in}
\OpFSCouple = \OpFSCouple_0 \bar \OpFSCouple \left( \frac{\mathbf{X}}{\ell} \right),  
 \hspace{0.2in}
 \OpFldDissp = \OpFldDissp_0 \bar{\OpFldDissp} = \frac{\mu}{\ell^2} \bar{\OpFldDissp}, 
 \hspace{0.2in}
\OpFSDissp = \OpFSDissp_0 \bar{\OpFSDissp}. 
\end{align}

The stochastic driving fields require some special considerations.  We find it convenient to express our scaling in the form
\begin{eqnarray}
\mathbf{g}_\subtxt{thm}\left(\mathbf{X}, s\right) = g_0 \mathbf{\bar{g}}_\subtxt{thm} = g_0 D_{-\bar{\OpFldDissp}} \bar{\xi}\left(\frac{\mathbf{X}}{\ell},\frac{s}{\tau_k}\right), \mbox{\hspace{0.7cm}}
\mathbf{F}_\subtxt{thm}\left(s\right) = F_0 \mathbf{\bar{F}}_\subtxt{thm} = F_0 D_{\bar{\OpFSDissp}} \bar{\eta}\left(\frac{s}{\tau_k}\right).
\mbox{\hspace{0.8cm}}
\end{eqnarray}
The $\bar{\xi}$ and $\bar{\eta}$ denote Gaussian random fields having mean zero and unit covariances
\begin{eqnarray}
\langle \bar{\xi}(\bar{\mathbf{X}},\bar{s}) \bar{\xi}^T(\bar{\mathbf{Y}},\bar{t}) \rangle = \delta(\bar{\mathbf{X}} - \bar{\mathbf{Y}}) \delta(\bar{s} - \bar{t}), \mbox{\hspace{0.7cm}}   
\langle \bar{\eta}(\bar{s}) \bar{\eta}^T(\bar{t}) \rangle =  \delta(\bar{s} - \bar{t}).
\mbox{\hspace{0.8cm}}
\end{eqnarray}
This allows for the characteristic strengths of the stochastic driving fields to be expressed as
\begin{align}
{g_0}^2 = \frac{ k_B T}{\ell^3 \tau_k} \OpFldDissp_0, &&
{F_0}^2 = \frac{ k_B T}{\tau_k} \OpFSDissp_0, &&
2 {D_A}{D_A}^T = A.
\end{align}
In our notation, $D_A$ denotes the square root of the operator $\frac{1}{2}A$ where we assume $A$ is positive semi-definite.

\subsection{Summary of Non-Dimensional Equations}
The SELM equations~\ref{equ_SELM_I_1}--~\ref{equ_SELM_I_3} can be expressed non-dimensionally and with operators treated to handle the constraints (see Section~\ref{sec_incorporating_incompressibility}) as
\begin{eqnarray}
\label{equ_SELM_I_nondim_1}
\frac{d\mathbf{\bar{p}}}{dt}  & = &\frac{1}{\delta} \bar{\OpFldDissp} \left(\mathbf{\bar{p}} - \kappa^{-1}  \bar{\OpSFCouple}[\mathbf{\bar{v}}]\right) + \alpha \bar{\OpSFCouple}[-\nabla_{\bar{X}}\bar{\Phi}(\mathbf{\bar{X}})] 
                         +\kappa^{-1}   \left(\nabla_{\bar{X}} \bar{\OpSFCouple}[\mathbf{\bar{v}}]\right)\cdot \mathbf{\bar{v}}
                          + \sqrt{\frac{1}{\delta}} \mathbf{\bar{g}}_\subtxt{thm} \\
\label{equ_SELM_I_nondim_2}
\frac{d\mathbf{\bar{v}}}{dt} & = & - \frac{1}{\epsilon}\bar{\OpFSDissp} C_1 (\mathbf{\bar{v}} - \mathbf{\bar{v}}_0) -    \kappa \alpha \nabla_{\bar{X}}\bar{\Phi}(\mathbf{\bar{X}})
                             + \sqrt{\frac{1}{\epsilon}}\sqrt{\kappa}\mathbf{\bar{F}}_\subtxt{thm} \\
\label{equ_SELM_I_nondim_3}
\frac{d\mathbf{\bar{X}}}{dt}  & = & \mathbf{\bar{v}}
\end{eqnarray}
where 
\begin{eqnarray}
C_1           = (I + \kappa^{-1} \bar{\OpFSCouple} \bar{\OpSFCouple}), \mbox{\hspace{0.5cm}} \mathbf{\bar{v}}_0 =  C_1^{-1}  \bar{\OpFSCouple} \mathbf{\bar{p}}, \mbox{\hspace{0.5cm}}
\langle \mathbf{\bar{g}}_\subtxt{thm}(s)\mathbf{\bar{g}}_\subtxt{thm}^{\RegAdjoint}(t) \rangle = -2\bar{\OpFldDissp}\hspace{0.06cm}\delta(t - s), \\ 
\langle \mathbf{\bar{F}}_\subtxt{thm}(s)\mathbf{\bar{F}}_\subtxt{thm}^{\RegAdjoint}(t) \rangle = 2\bar{\OpFSDissp}\hspace{0.06cm}\delta(t - s), \mbox{\hspace{0.5cm}} \langle \mathbf{\bar{g}}_\subtxt{thm}(s)\mathbf{\bar{F}}_\subtxt{thm}^{\RegAdjoint}(t) \rangle =  0. \mbox{\hspace{2.3cm}}
\label{equ_SELM_I_6_bar}
\end{eqnarray}
The characteristic scales are defined in Section~\ref{sec_Char_Scales}.
The incompressibility and microstructure constraints are handled as discussed in Section~\ref{sec_incorporating_incompressibility}.
We assume throughout that $\alpha \kappa = 1$ to ensure that the conservative forces remain comparable to other influences on the microstructure.
In Section~\ref{sec_derivation_rapid_fluid_relax} 
 we will consider $\delta \to 0$, but only after the limits $\epsilon \to 0$ followed by the limit $\kappa^{-1}\to0$.

\section{Stochastic Reduction}
\label{sec_reduction}

We derive reduced equations for the fluid-structure dynamics when 
(i) the coupling forces that transfer momentum between the fluid and microstructures is strong
($\epsilon \rightarrow 0$), 
(ii) the mass of the microstructures is small relative to the displaced mass of the fluid ($\kappa^{-1} \rightarrow 0$), 
and (iii) the response to stresses results in hydrodynamics that relax rapidly to 
a quasi-steady-state relative to the motions of the microstructure ($\delta \rightarrow 0$).
We present the leading order behavior in the asymptotic parameter, and in some cases also 
the first order corrections.  We first give a summary of our main results in this section, 
describe our mathematical analysis in Section~\ref{sec_general_reduction_procedure}, and give  detailed derivations of these results in Section~\ref{sec_derivation_reduction}.  We remark that to simplify the notation we have dropped annotation of variables in previous sections but the equations in this section are dimensional.

\subsection{Limit of Strong Coupling: Summary of Reduced Equations}
\label{section_summary_strong_coupling}
In this section and the following sections, we summarize the dimensional equations that result from our reductions performed on the non-dimensional equations.
We consider the limit $\epsilon \rightarrow 0$ of equations~\ref{equ_SELM_I_1}--\ref{equ_SELM_I_3}  where the fluid and the microstructures 
become strongly coupled and momentum is exchanged rapidly. 
In terms of the physical parameters this regime occurs when the momentum coupling parameter satisfies 
$\Upsilon_0 \gg \sqrt{m_0 k_B{T}/\ell^2}$.
We obtain the effective inertial dynamics 
\begin{eqnarray}
\label{equ_reduced_sc_p}
\frac{d\mathbf{p}}{dt}  & = & \rho^{-1} \OpFldDissp \left(\mathbf{p} 
                          - \OpSFCouple[m\mathbf{v}_0]\right) 
			  - \OpSFCouple \nabla_{\mathbf{X}}\Phi(\mathbf{X})  
                          + \left(\nabla_{\mathbf{X}} 
 \OpSFCouple[m\mathbf{v}_0]\right)\cdot \mathbf{v}_0 \nonumber \\ 
                        & + & k_B T \nabla_{\mathbf{X}} \OpSFCouple : C_1^{-1}                 
				+\lambda          
                          + \mathbf{g}_\subtxt{thm} 
				+  \epsilon\tilde{\theta}_p
\\
\label{equ_reduced_sc_v}
\frac{d\mathbf{X}}{dt}  & = & \mathbf{v}_0
				+  \epsilon\tilde{\theta}_X.
\end{eqnarray}
In the notation, the double dot-product should be interpreted as $\left(\nabla_{\mathbf{X}} \OpSFCouple : C_1^{-1}\right)_i = \partial_{X_k} \OpSFCouple_{ij} (C_1^{-1})_{jk}$.
The $\mb{v}_0$ denotes the effective velocity of the microstructures to leading order:
\begin{eqnarray}
\label{equ_v_0_finite_kappa}
\mathbf{v}_0  = \rho^{-1} C_1^{-1} \OpFSCouple \mathbf{p}, \mbox{\hspace{1cm}} \text{ with  }
C_1             =  I + {\rho}^{-1}{m} \OpFSCouple \OpSFCouple. 
\end{eqnarray}
The thermal fluctuations are taken into account through the Gaussian stochastic driving field $\mathbf{g}_\subtxt{thm}$ with mean zero and covariance
\begin{eqnarray}
\langle \mathbf{g}_\subtxt{thm}(s)\mathbf{g}_\subtxt{thm}^{\RegAdjoint}(t) \rangle 
& = & -\left(2k_B{T}\right) \OpFldDissp \hspace{0.06cm}\delta(t - s). 
\end{eqnarray}
The $\theta = \epsilon \tilde{\theta}$ terms represent higher-order terms at the next order in $\epsilon$.  For the strong coupling limit they capture the leading order correction from slip effects between the fluid and structures in this regime.  These terms could also be useful  in capturing the permeation of fluid through a structure within the fluid such as a porous 
membrane~\cite{AtzbergerSwelling2014}.  
In particular, the $\theta_X = \epsilon \tilde{\theta}_X$ results from the slip of a microstructure relative to the background fluid flow corresponding to a non-zero $\epsilon$ (non-infinite $\Upsilon$).  
These terms capture to some approximation the transient responses of a microstructure to an applied body force. 
Similar terms have been found for inertial equations derived using other approaches~\cite{Kim2007a, Donev2012}.   For detailed expressions for these terms found using our systematic reductions see Section~\ref{section_weak_slip}.    

\subsubsection{Limit of Strong Coupling: Reduced Equations in Terms of the Effective Fluid Velocity $\mb{u}$}
\label{section_fluid_equ}
The effective fluid velocity field $\mb{\bar{u}}$ is obtained from the total momentum field  by 
\begin{eqnarray}
\label{equ_fluid_equ_total_mom}
\mb{\bar{u}} = \rho^{-1}\left(\mb{p} - \OpSFCouple[m(\mathbf{v}_0 + \epsilon\tilde{\theta}_X) ] \right).
\end{eqnarray}
The $\mathbf{v}_0$ is given in Section~\ref{section_summary_strong_coupling} and $\theta_X = \epsilon\tilde{\theta}_X$ is given in Section~\ref{section_weak_slip}.
The effective fluid-structure dynamics  of~\ref{equ_reduced_sc_p}--\ref{equ_reduced_sc_v} are
\begin{align}
\label{equ_u_bar_inertial_fld}
\rho \frac{d \mathbf{\bar{u}}}{dt} 
& =  \OpFldDissp\bar{\mathbf{u}}
					+k_B T \nabla_{\mathbf{X}} \OpSFCouple  : C_1^{-1} 
					-\OpSFCouple \frac{d}{dt} [m\mathbf{v}_0]
		- \OpSFCouple \nabla_{\mathbf{X}}\Phi(\mathbf{X}) 
				+ \epsilon \tilde \theta_{\bar{u}}
                         + \lambda + \mathbf{g}_\subtxt{thm}  \\ 
\label{equ_u_bar_particle_X}	
\frac{d\mathbf{X}}{dt} 
&= \mathbf{v}_0 + \epsilon\tilde{\theta}_X = \OpFSCouple \bar{\mathbf{u}} + \epsilon C_1 \tilde{\theta}_X \\
\langle \mathbf{g}_\subtxt{thm}(s)\mathbf{g}_\subtxt{thm}^{\RegAdjoint}(t) \rangle 
&= -\left(2k_B{T}\right) \OpFldDissp \hspace{0.06cm}\delta(t - s). 
\end{align}
The effective velocity  of the microstructures   $\mb{v}_0$  can be expressed to order $\epsilon$  as
\begin{eqnarray}
\label{equ_v_0_finite_kappa2}
\mathbf{v}_0 & = & \OpFSCouple \mathbf{\bar{u}} +\epsilon \rho^{-1} m \OpFSCouple \OpSFCouple \tilde{\theta}_X. 
\end{eqnarray}
The $\epsilon$-order correction term for $\frac{d \mathbf{\bar u}}{dt}$ can be expressed as
\begin{align}
\tilde \theta_{\bar{u}} = \tilde \theta_p +  \rho^{-1} \OpFldDissp\left[\Lambda ( m \tilde \theta_X) \right] -  \frac{d}{dt}  \left[  \OpSFCouple(m \tilde{\theta}_X)  \right]- \nabla_{\mathbf{X}} \OpSFCouple(m \mathbf{v_0}) \cdot \tilde \theta_X  .
\end{align}
The equations in the presented form were obtained by using 
\begin{align}
				&\left(\nabla_{\mathbf{X}} \OpSFCouple[m\mathbf{v}_0]\right)\cdot \mathbf{v}_0  - \frac{d}{dt}( \OpSFCouple[m \mathbf{v}_0])  \\
&= \left(\nabla_{\mathbf{X}} \OpSFCouple[m\mathbf{v}_0]\right)\cdot \mathbf{v}_0  
-\left(\nabla_{\mathbf{X}} \OpSFCouple[m\mathbf{v}_0]\right)\cdot \mathbf{v}_0  
-\OpSFCouple \frac{d}{dt} [m\mathbf{v}_0]
= - \OpSFCouple \frac{d}{dt} [m\mathbf{v}_0].
\end{align}
The effective fluid equations~\ref{equ_u_bar_inertial_fld} account for changes in momentum arising from the microstructure forces, microstructure inertial effects, hydrodynamic shear stresses within the fluid, and thermal fluctuations.

Another interesting feature of the analysis is that a direct approach attempting to use the fluid velocity field during the 
reduction presents difficulties arising from the increasingly rapid momentum exchange between the fluid and microstructures 
as the coupling becomes strong.  To 
avoid these rapid dynamics, we reformulated the system in terms of the total momentum density 
field $\mathbf{p}$ in Section~\ref{sec_reform_total_mom} as introduced in~\cite{AtzbergerSELM2011}.  This difficulty indicates that the instantaneous fluid velocity does not behave like a ``slow 
variable'' in the strong coupling limit.  However, when using the total momentum field as the ``slow variable'' the analysis readily 
proceeds.  An interesting outcome, which we initially found a bit curious and counter-intuitive, is that one can again express the 
reduced equations in terms of the fluid velocity field $\mb{\bar{u}}$.  What should be realized is that the interpretation 
of this field takes on a slightly different meaning and should be thought of intuitively as an effectively time-averaged 
fluid velocity over some small time-scale (instead of the instantaneous fluid velocity).  The mathematical analysis 
systematically handles this to provide a set of equations for an effectively ``renormalized'' fluid velocity.
We can view our transformation to the total momentum density field and our systematic reduction analysis 
as a precise way to perform such an averaging.
 
\subsection{Limit of Negligible Excess Mass: Summary of Reduced Equations}
When the microstructure excess mass is small relative to the local 
displaced fluid, the equations can be further reduced in 
the limit $\kappa^{-1} \rightarrow 0$.  In terms of the physical parameters,
we approach this limit when the excess mass parameter satisfies $m \ll \rho \ell^3$.  
We assume that $\epsilon$ was already considered small and thus start from equations~\ref{equ_u_bar_inertial_fld}--\ref{equ_u_bar_particle_X}.
This results in the reduced fluid-structure equations
\begin{eqnarray}
\label{equ_u_bar_m_zero_fld}
\rho\frac{d\mb{u}}{dt}  & = & \OpFldDissp \mb{u}   
                          - \OpSFCouple \nabla_{\mb{X}}\Phi(\mb{X}) 
                          + k_B T \nabla_{\mb{X}} \cdot \OpSFCouple 
                          + \lambda
				+\epsilon\tilde{\theta}_u
                          + \mb{f}_\subtxt{thm} \\
\label{equ_X_m_zero_fld}
\frac{d\mb{X}}{dt}  & = & \OpFSCouple \mb{u} + \epsilon\tilde{\theta}_X \\
\label{equ_f_thm_m_zero_fld}
\left \langle \mb{f}_\subtxt{thm}(t) \mb{f}_\subtxt{thm}(s)^T \right \rangle & = & 
-2 k_B{T} \OpFldDissp \delta(t - s).
\end{eqnarray}
We derive these these results in Section~\ref{sec_derivation_small_excess_mass}.
Notice our result for the $\theta = \epsilon\tilde{\theta}$ terms  given in Section~\ref{section_weak_slip} can be simplified in consideration of taking  $\kappa^{-1} \to 0$.

In this limit, the microstructure inertial terms disappear in equations~\ref{equ_u_bar_inertial_fld}--\ref{equ_u_bar_particle_X}.  This physical
regime is very similar to what is treated in the \textit{Stochastic Immersed Boundary Method} (SIB) in~\cite{Peskin2002,Atzberger2007a}.  
However, the systematic reduction analysis we perform here shows that there is an important drift term $ k_B T \nabla_{\mb{X}} \cdot \OpSFCouple$ that 
does not appear
in the original SIB formulation~\cite{Atzberger2007a}.  
The term $k_B T \nabla_{\mathbf{X}} \OpSFCouple  : C_1^{-1}$ in equation~\ref{equ_u_bar_inertial_fld} when $m = 0$ yields the drift $k_B T \nabla_{\mb{X}} \cdot \OpSFCouple$ in equation~\ref{equ_u_bar_m_zero_fld}.  The drift term arises
from the generalized coordinates used for the fluid and microstructures that have configuration 
and momentum variables that are non-conjugate in the Hamiltonian sense.  Consequently, the Liouville theorem for the invariant distribution under the phase-space dynamics must be generalized to include a metric factor which manifests as a drift term in the stochastic dynamics, see~\cite{AtzbergerSELM2011}.   

In the zero excess mass limit, $m = 0$, there is no longer a distinction between the dynamics of the total momentum field and the usual
fluid velocity equations 
to leading order in $\epsilon$ since $\mb{p} = \rho\mb{u}$.  Our results suggest that approaches such as the 
Immersed Boundary Method~\cite{Peskin2002} which treat this regime should be interpreted as assuming that the 
microstructure is effectively density matched with the fluid.  Furthermore, any inertial effects of the 
microstructures arise from the entrained motions of the fluid parcels overlapping with the
kernel functions used in the operators for velocity averaging and force-spreading~\cite{Peskin2002,AtzbergerSELM2011,Atzberger2007a}.  
To model additional inertial effects of microstructures within IB, some recent extensions to IB have been 
introduced in~\cite{Kim2007a, Donev2012}.

\subsection{Limit of Rapid Hydrodynamic Relaxation: Summary of Reduced Equations}
We consider the regime where the hydrodynamics rapidly equilibrate to a 
quasi-steady state in response to body forces.  In terms of
the physical parameters this occurs when the fluid momentum spreads rapidly relative to
the microstructure motion $\rho \ell^2/\mu \ll \ell/U$ or 
equivalently when the viscosity is large in the sense $\mu \gg \rho \ell U$.    
We consider the case when the limit $\epsilon \to 0$ has already been taken 
so that slip corrections are negligible and $\kappa^{-1} \to 0$ so the structures contribute negligible excess mass. Thus we begin with \ref{equ_u_bar_m_zero_fld}--\ref{equ_f_thm_m_zero_fld}.  In this regime, we obtain the reduced equations
\begin{eqnarray}
\frac{d\mathbf{X}}{dt} &=&
		 H_{\subtxt{SELM}} [-\nabla_{\mb{X}}\Phi(\mb{X})]
		+ (\nabla_\mathbf{X} \cdot H_{\subtxt{SELM}}) k_B T 
		+\mathbf{h}_{\subtxt{thm}}\\
H_{\subtxt{SELM}} &=&  \OpFSCouple (\proj \OpFldDissp)^{-1} \OpSFCouple, \mbox{\hspace{0.4cm}} 
\langle \mathbf{h}_{\subtxt{thm}}(s), \mathbf{h}^T_{\subtxt{thm}}(t) \rangle = 2 k_B T H_{\subtxt{SELM}} \delta(t-s).
\end{eqnarray}
In the notation, $\OpFldDissp = \mu \Delta$.  In this regime the fluid degrees of freedom are eliminated entirely and replaced by the 
effective hydrodynamic coupling tensor $H_{\subtxt{SELM}}$.  Interestingly, the drift term arising 
from the generalized fluid-structure coordinates manifests as the term 
$(\nabla_\mathbf{X} \cdot H_{\subtxt{SELM}}) k_B T$.  This drift-divergence term is essential for 
the microstructure dynamics to have invariant the Gibbs-Boltzmann distribution with detailed balance~\cite{AtzbergerSELM2011}.
We derive these results in Section~\ref{sec_derivation_rapid_fluid_relax}.

\section{Stochastic Reduction Method: Singular Perturbation of Backward Kolmogorov Equations}
\label{sec_general_reduction_procedure}

We derive the reduced equations using a formal method based 
on a singular perturbation analysis  
of the Backward-Kolmogorov 
Equations (BKE)~\cite{Kramer2004,Majda2003,Kurtz1973,Papanicolaou1976}. 
The BKE are given by  
\begin{eqnarray}
\label{equ_BKE}
\frac{\partial \bkFunc}{\partial t} = \mathcal{A} \bkFunc, \mbox{\hspace{1cm}} \bkFunc(0,\mb{z}) = \avgFunc(\mb{z}).
\end{eqnarray}
The
$\bkFunc(t,\mb{z}) = E^{\mb{z}}[\avgFunc(\mb{Z}(t))]$
with $\mb{Z}(t) = (\mathbf{X}(t),\mathbf{p}(t),\mathbf{v}(t))$~\cite{Oksendal2000}.
The $E^{\mb{z}}[\cdots]$ denotes 
taking expectation when the stochastic process starts with $\mb{Z}(0) = \mb{z}$.
The $\mathcal{A}$ is the infinitesimal generator of the stochastic process $\mb{Z}(t)$.  
The $\avgFunc$ is assumed to be a $C^2$ smooth function 
with compact support.  An important connection is that the 
statistics of the stochastic process are determined by the 
expectations taken over the class of functions $f$.  This provides a mapping between 
the infinitesimal generator $\mathcal{A}$ that appears in the BKE and the 
underlying stochastic process $\mb{Z}(t)$.
In particular, consider the stochastic process satisfying 
\begin{eqnarray}
\label{equ_SDE_dZ}
d\mb{Z}(t) = \mb{a}(\mb{Z}(t)) dt + \mb{b}(\mb{Z}(t)) d\mb{W}_t.  
\end{eqnarray}
The Stochastic Differential Equation (SDE) is to be given the the Ito interpretation~\cite{Oksendal2000}.
The corresponding infinitesimal generator is 
\begin{eqnarray}
\mathcal{A} & = & \mb{a}\cdot \frac{\partial}{\partial \mb{z}} + \frac{1}{2} \mb{b}\mb{b}^T : \frac{\partial^2}{\partial \mb{z}^2}.
\end{eqnarray}
We shall use this to obtain the reduction by performing a singular 
perturbation analysis of the BKE to yield a limiting form for the 
infinitesimal generator 
$\mathcal{\tilde{A}} = \mb{\tilde{a}}\cdot {\partial}/{\partial \mb{z}} + \frac{1}{2} \mb{\tilde{b}}\mb{\tilde{b}}^T : {\partial^2}/{\partial \mb{z}^2}$.
This determines a reduced stochastic process $\mb{\tilde{Z}}(t)$ satisfying 
$d\mb{\tilde{Z}}(t) =  \mb{\tilde{a}}(\mb{\tilde{Z}}(t)) dt + \mb{\tilde{b}}(\mb{\tilde{Z}}(t)) d\mb{\tilde{W}}_t$
which approximates the full stochastic dynamics of equation~\ref{equ_SDE_dZ}. 
We remark that a distinct 
advantage of using the BKE over the Fokker-Planck Equations (FPE) is that our 
approximation will not be required to satisfy additional constraints, such as ensuring
the equation conserves the total probability density.  The adjoint of an approximated FPE
differential operator is not always a valid infinitesimal generator.  In contrast when
making approximations of the BKE, the obtained second order differential operator 
has the plausible form for an infinitesimal generator \cite{Oksendal2000}.   

To obtain a specific limiting regime, we require that terms be identified that split
the dynamics of the infinitesimal generator into ``slow'' and ``fast'' parts 
\begin{equation}
\mathcal{A}_\epsilon = L_{slow}  + L_{fast}.
\end{equation}
As the notation suggests, the splitting is meant to separate the 
degrees of freedom of the system into two classes $\mb{z} = (\mb{z}_s, \mb{z}_f)$.  
The $\mb{z}_s$ are those 
that exhibit relatively ``slow"  temporal dynamics.  The $\mb{z}_f$ are those degrees of freedom that exhibit 
relatively ``fast" temporal dynamics.  The $L_{slow}$ contains only the terms of the infinitesimal generator
that involves derivatives with respect to the ``slow" degrees of freedom.  Similarly, 
the $L_{fast}$ contains only the terms governing the ``fast" degrees of freedom. 
In order for such a splitting to exist, we require $\mathcal{A}_\epsilon$ to have no terms 
with mixed derivatives in the slow and fast degrees of freedom of the form $\partial^2/\partial z_s\partial z_f$ .
 These notions are defined
more precisely for the SELM dynamics in Section~\ref{sec_derivation_reduction}.

This splitting provides a useful stationary conditional probability distribution $\Psi(\mb{z}_f | \mb{z}_s)$ for 
the ``fast" degrees of freedom $\mb{Z}_f(t)$ when evolving under dynamics with the 
``slow" degrees of freedom held fixed $\mb{Z}_s = \mb{z}_s$.
This is given by solving
the steady-state FPE for the fast degrees of freedom which can be expressed using 
the adjoint of the generator as a solution of
\begin{eqnarray}
\label{equ_Psi_steady}
L_{2}^* \Psi =  0, \mbox{\hspace{1cm}} \int \Psi d\mb{z}_f =  1.
\end{eqnarray}
More precisely, we shall consider in our analysis for the ``fast'' degrees of freedom 
generators of the general form
\begin{eqnarray}
L_{fast} & = & \frac{1}{\epsilon} \left( L_2 + \epsilon \tilde{L}_2 \right).
\end{eqnarray}
The $L_2$ represents the leading order contribution to the generator $L_{fast}$
and is used to determine the invariant distribution $\Psi$.  Often we have  
$\tilde{L}_2 = 0$, but as we shall discuss, for many cases of interest this 
term is non-zero making an interesting contribution to the 
reduced effective stochastic dynamics.

For the ``slow'' degrees of freedom, we find it convenient to split 
the generator as 
\begin{eqnarray}
L_{slow} = \bar{L}_{1} + L_1, \mbox{\hspace{0.8cm}}
\bar{L}_{1} =  \int \Psi(\mb{z}_f | \mb{z}_s) L_{slow} d\mb{z}_f, \mbox{\hspace{0.8cm}}
L_1         = L_{slow} - \bar{L}_{1}.
\end{eqnarray}
This splitting ensures that $L_1$ generates a stochastic process having mean zero.  
As we shall discuss, if $\bar{L}_{1}$ is a non-zero operator then it captures the
leading order dynamics.  The $L_1$ then contributes at the next order.  
These conventions allow for the 
infinitesimal generator to be expressed as
\begin{eqnarray}
\label{equ_final_A_split}
\label{equ_L_epsilon}
\mathcal{A}_\epsilon & = & \bar{L}_{1}  + \epsilon L_{\epsilon},  \mbox{\hspace{0.8cm}}
L_\epsilon = \f{1}{\epsilon}\left(L_1 + \tilde{L}_2\right)+ \f{1}{\epsilon^2} L_2. 
\end{eqnarray} 
As we shall show, the operator $L_{\epsilon}$ contributes 
effectively as order one in the limit $\epsilon \rightarrow 0$
and hence the scaling and notation chosen.

To make this more precise, we make the ansatz that $u$ can be expanded as
\begin{eqnarray}
\label{equ_expansion}
u(\mb{z},t) & = & u_0(\mb{z},t) + u_1(\mb{z},t)\epsilon + u_2(\mb{z},t)\epsilon^2 \cdots + u_n(\mb{z},t)\epsilon^n + \cdots.
\end{eqnarray}
We shall seek ultimately a partial differential equation (BKE) for the first two orders
\begin{eqnarray}
\bar{u}(\mb{z}_s,t) & = & u_0(\mb{z}_s,t) + \bar{u}_1(\mb{z}_s,t)\epsilon
\end{eqnarray}
where $\bar{u}_1(\mb{z}_s,t) = \int \Psi(\mb{z}_f | \mb{z}_s) u_1(\mb{z},t) d\mb{z}_f$.
By comparing orders when plugging equation~\ref{equ_expansion} into~\ref{equ_BKE} and using~\ref{equ_final_A_split} we obtain
\begin{eqnarray}
O\left(\epsilon^{-1}\right) : L_2 u_0 & = & 0 \\ 
O\left(1\right) : \frac{\partial u_0}{\partial t} & = & \bar{L}_1 u_0 + L_1 u_0 + \tilde{L}_2 u_0 + L_2 u_1 \\
O\left(\epsilon\right) : \frac{\partial u_1}{\partial t} & = & \bar{L}_1 u_1 + L_1 u_1 + \tilde{L}_2 u_1 + L_2 u_2. 
\end{eqnarray}
We assume throughout that the stochastic process generated by $L_2$ is ergodic on the space of $\mb{z}_f$ so 
that $\mbox{dim} \hspace{0.1cm} \mbox{ker} \{L_2^*\} = 1$.  The order $O\left(\epsilon^{-1}\right)$ can be 
interpreted as the steady-state of the Backward-Kolmogorov equation of a stochastic process $\mb{\hat{Z}}_f(t)$ generated by $L_2$.
This suggests that $u_0(\mb{z}) = \lim_{t \rightarrow \infty}E^{\mb{z}}\left\lbrack f(\mb{\hat{Z}}_f(t) ) \right\rbrack = E^{\mb{z}_s}\left\lbrack f(\mb{\hat{Z}}_f) \right\rbrack = u_0(\mb{z}_s)$, where $\mb{\hat{Z}}_f(t)$ is the process started with $\mb{\hat{Z}}_f(t) = \mb{z}_f$.
By ergodicity the long-term behavior of $\mb{\hat{Z}}_f(t)$ would be independent of the initial condition 
and the latter expectation is to be taken with 
respect to $\Psi$ satisfying equation~\ref{equ_Psi_steady}.  This gives that $u_0 = u_0(\mb{z}_s)$ with the only dependence on 
$\mb{z}_s$.  Throughout we take $u_0$ only depending on $\mb{z}_s$ which ensures the 
order $O\left(\epsilon^{-1}\right)$ is always satisfied since $L_2$ only involves derivatives with 
respect to $\mb{z}_f$.    The order $O\left(1\right)$ can be 
used to solve for $u_1$ in terms of $u_0$ by
\begin{eqnarray}
\label{equ_L2_u1}
L_2 u_1 & = & \frac{\partial u_0}{\partial t} - \bar{L}_1 u_0 - L_1 u_0.
\end{eqnarray}
We used that $\tilde{L}_2 u_0 = 0$ since $u_0 = u_0(\mb{z}_s)$ and $\tilde{L}_2$ only involves derivatives in $\mb{z}_f$.
The solvability of equation~\ref{equ_L2_u1} requires the right-hand side of the equation be in the range of the operator $L_2$.  A well known condition
for this is that the right-hand side be orthogonal to all elements of the null-space of $L_2^*$.  In other 
words, the $\overline{\mbox{range}\{L_2\}} = \overline{\mbox{ker}\{L_2^*\}^{\perp}}$, where ${\perp}$ denotes
the orthogonal compliment of a set under the standard $L^2$-inner product, see~\cite{Reed1980}.
By our ergodicity assumption the kernel only has one dimension and the solvability can be represented by the condition 
\begin{eqnarray}
\label{equ_u1_solvability}
\left(\int \Psi \left(\frac{\partial}{\partial t} - \bar{L}_1  - L_1\right) d \mb{z}_f\right) u_0 = 0.
\end{eqnarray}
The $\Psi$ is the stationary probability density satisfying equation~\ref{equ_Psi_steady}.
This yields the BKE for the leading order 
\begin{eqnarray}
\label{equ_reduction_u0}
\frac{\partial u_0}{\partial t} = \bar{L}_1 u_0.
\end{eqnarray}
This follows since by definition $\int \Psi L_1 d\mb{z}_f = 0$ and the $\bar{L}_1 = \int \Psi \bar{L}_1 d\mb{z}_f$ 
since it has already been averaged with respect to the probability distribution.  The condition for the 
existence of the solution $u_1$ in the asymptotic expansion expressed in equation~\ref{equ_L2_u1} 
provides the equation~\ref{equ_reduction_u0} for the 
leading order $u_0$.  Using equation~\ref{equ_L2_u1}, the order $u_1$ can be expressed as 
\begin{eqnarray}
u_1 & = & L_2^{-1} \left(\frac{\partial u_0}{\partial t} - (\bar{L}_1 + L_1)u_0 \right) = -L_2^{-1}L_1 u_0.
\end{eqnarray}
The final expression comes from the relationship of the partial derivative $\partial u_0/ \partial t$ and the operator 
$\bar{L}_1$ given by equation~\ref{equ_reduction_u0}.  Now at the order $O(\epsilon)$ a 
very similar argument can be made to ensure the solvability of $u_2$.  This yields 
\begin{eqnarray}
\label{eqn_reduction_u1}
\frac{\partial \bar{u}_1}{\partial t} & = & - \left(\int \Psi \left(\bar{L}_1 + L_1 + \tilde{L}_2\right)L_2^{-1} L_1 d\mb{z}_f \right) u_0
\end{eqnarray}
where $\bar{u}_1(\mb{z}_s,t) = \int \Psi(\mb{z}_f | \mb{z}_s) u_1(\mb{z},t) d\mb{z}_f$.
This provides a closed set of differential equations for the first two orders $u_0$, $\bar{u}_1$ approximating the 
solution of the BKE in the $\epsilon \rightarrow 0$ limit, see equations~\ref{equ_reduction_u0} 
and~\ref{eqn_reduction_u1}.  

It is convenient to express this approximation by deriving a set of closed equations for 
$\bar{u} = u_0 + \epsilon \bar{u}_1$.  We have that 
\begin{eqnarray}
\label{equ_bar_u_u0}
\frac{\partial \bar{u}}{\partial t}  = \bar{L}_1 u_0 + \epsilon \left(-\int \Psi \left(\bar{L}_1 + L_1 + \tilde{L}_2 \right)L_2^{-1}L_1 d\mb{z}_f \right) u_0.
\end{eqnarray}
To express this in terms of $\bar{u}$ it is useful to notice that 
\begin{eqnarray}
\bar{u} = \left(\mathcal{I} - \epsilon \int \Psi L_2^{-1} L_1 d\mb{z}_f \right) u_0.
\end{eqnarray}
By inverting this operator and expanding to leading orders in $\epsilon$ we have 
\begin{eqnarray}
u_0 =  \left(\mathcal{I} + \epsilon \int \Psi L_2^{-1} L_1 d\mb{z}_f + \epsilon^2(\cdots) + \cdots\right) \bar{u}
\end{eqnarray}
By neglecting orders greater than $\epsilon$ we have 
\begin{eqnarray}
\label{equ_L1_u0}
\bar{L}_1 u_0 & = & \bar{L}_1 \bar{u} + \epsilon \bar{L}_1 \int \Psi L_2^{-1} L_1 d\mb{z}_f \bar{u}.
\end{eqnarray}
This gives the final set of closed reduced equations  
\begin{eqnarray}
\label{equ_du_reduc}
\label{equ_L_0}
\frac{\partial \bar{u}}{\partial t} & = & \left( \bar{L}_1 + \epsilon \bar{L}_0\right) \bar{u}, \mbox{\hspace{0.8cm}}
\bar{L}_0 = -\int \Psi \left( L_1 + \tilde{L}_2\right) L_2^{-1} L_1 d\mb{z}_f .
\end{eqnarray}
This follows by using equation~\ref{equ_L1_u0} in equation~\ref{equ_bar_u_u0} and canceling common terms.  The operator $\bar{L}_0$ is obtained by collecting after cancellations all of the remaining terms of order $\epsilon$.
This derivation provides a unified expression consistent with the formal methods used 
in~\cite{Majda2004a,Kramer2004,Majda2003} and the rigorous results obtained in~\cite{Kurtz1973,Papanicolaou1976}.
This provides a BKE with generator $\tilde{\mathcal{A}} = \bar{L}_1 + \epsilon \bar{L}_0$ approximating 
the full BKE given in equation~\ref{equ_BKE}.  
Interestingly, the term $L_\epsilon$ of equation~\ref{equ_L_epsilon} is approximated in the 
final set of equations by $\bar{L}_0$ which contributes only as order one in $\epsilon$,
``$L_\epsilon \rightarrow \bar{L}_0$.''  The operator $\tilde{\mathcal{A}}$ provides the 
infinitesimal generator for the reduced stochastic process $\mb{\tilde{Z}}(t)$ approximating 
the full stochastic process $\mb{Z}(t)$ given by equation~\ref{equ_SDE_dZ}.  We remark that the full 
stochastic process can be viewed as approximating the reduced stochastic process in the sense of "weak convergence" for the observable $u$ defined after equation~\ref{equ_BKE}.
The 
equations~\ref{equ_du_reduc} 
establish
our systematic reduction procedure to approximate the full stochastic dynamics.

\section{Derivation of the Non-Dimensional Reduced Equations}
\label{sec_derivation_reduction}

The details are now presented for the derivation of the various reduced equations in the limits of (i) strong coupling, (ii) small body excess mass, and 
(iii) rapid hydrodynamic relaxation.  The equations are reduced by decomposing the infinitesimal generator into ``fast'' and ``slow'' parts using 
the singular perturbation analysis presented in Section~\ref{sec_general_reduction_procedure}.  A central challenge to compute the effective infinitesimal generator 
is to invert the ``fast'' operator $L_2$.  We give a general method for Sturm-Liouville operators in Appendix~\ref{sec_sl}.  We identify 
each regime precisely using our dimensional analysis and then provide appropriate decompositions to obtain an explicit expression 
for the reduced equations.    

\subsection{Limit of Strong Coupling: Derivation of Reduced Equations}
\label{sec_derivation_strong_coupling}

We derive the reduced equations in the regime when the coupling for the momentum exchange 
between the fluid and the microstructures becomes strong $\epsilon \rightarrow 0$.  

\subsubsection{Splitting of the Infinitesimal Generator into Slow and Fast Parts}  

To avoid clutter, we will remove the bars from the nondimensional equation~\ref{equ_SELM_I_nondim_2} and use this convention throughout our derivation.
To handle the infinitesimal generator in this regime, it is very useful to 
make the change of variable in the velocity 
$\mathbf{\tilde{v}} = \mathbf{v}-\mathbf{v_0}$.  Specifically, we define 
\begin{eqnarray}
\label{equ_v0}
\label{equ_C_1}
\label{equ_C}
\mathbf{v}_0 & = & C_1^{-1}  {\OpFSCouple} \mathbf{{p}}, \mbox{\hspace{0.8cm}} 
C_1   = (I + \kappa^{-1} {\OpFSCouple} {\OpSFCouple}).
\end{eqnarray}
This serves to center the terms up to order $\epsilon$ the 
equation~\ref{equ_SELM_I_nondim_2} and allows for the equations 
to be put into the convenient form
\begin{eqnarray}
\label{equ_SELM_I_nondim_1_cen}
\frac{d\mathbf{{p}}}{dt}  & = & \frac{1}{\delta} {\OpFldDissp} \left(\mathbf{{p}} - \kappa^{-1} {\OpSFCouple}[\mathbf{v}]\right) + \alpha{\OpSFCouple}[-\nabla_{{X}}{\Phi}(\mathbf{{X}})] 
                         + \kappa^{-1} \left(\nabla_{{X}} {\OpSFCouple}[\mathbf{v}]\right)\cdot (\mathbf{v})
                          +\sqrt{\frac{1}{\delta}}  \mathbf{{g}}_\subtxt{thm} \\
\label{equ_SELM_I_nondim_2_cen}
\frac{d\mathbf{\tilde{v}}}{dt} & = & 
		\frac{d\mathbf{{v}}}{dt} - \frac{d\mathbf{{v_0}}}{dt} 
		= - \frac{1}{\epsilon}{\OpFSDissp} C_1 \mathbf{\tilde{v}} 
                        + \sqrt{\frac{1}{\epsilon}} \sqrt{\kappa} \mathbf{{F}}_\subtxt{thm} 
			 -  \alpha \kappa \nabla_{{X}}{\Phi}(\mathbf{{X}}) - \nabla_{\mathbf{X},\mathbf{p}} \mathbf{v_0} \cdot \frac{d}{dt}(\mathbf{X},\mathbf{p})\\
\label{equ_SELM_I_nondim_3_cen}
\frac{d\mathbf{{X}}}{dt}  & = & \mathbf{v}.
\end{eqnarray}
The last term in~\ref{equ_SELM_I_nondim_2_cen} results from a use of the chain rule when differentiating $\mathbf{v}_0$.
The infinitesimal generator of this fluid-structure system is split into the parts
\begin{align}
\mathcal{A} = &  L_{slow} +  L_{fast} \\
\label{L_slow}
L_{slow} = & \bigg[ \frac{1}{\delta} \OpFldDissp \left(\mathbf{p} - \kappa^{-1} \OpSFCouple[\mathbf{v}]\right) + \alpha \OpSFCouple[-\nabla_{\mathbf{X}}\Phi(\mathbf{X})] 
                         + \kappa^{-1} \left(\nabla_{\mathbf{X}} \OpSFCouple[\mathbf{v}]\right)\cdot \mathbf{v} \bigg] \cdot \nabla_{\mathbf{p}} \\
\nonumber
					 -& \frac{1}{\delta} \mathcal{L} : \nabla_{\mathbf{p}}^2
						+\mathbf{v} \cdot \nabla_\mathbf{X} \\
\label{L_fast}
L_{fast} = &  \frac{1}{\epsilon}\left[ (- \OpFSDissp C_1 \mathbf{\tilde{v}} ) \cdot \nabla_{\mathbf{\tilde{v}}} 
		 + \kappa \OpFSDissp : \nabla_{\mathbf{\tilde{v}}}^2 \right] - \alpha\kappa \nabla_{\mathbf{X}} \Phi(\mathbf{X})\cdot \nabla_{\mathbf{\tilde{v}}}  -  
\mathcal{A}_{\mathbf{\tilde v}} \left[\nabla_{\mathbf{X},\mathbf{p}} \mathbf{v_0} \cdot \frac{d}{dt}(\mathbf{X},\mathbf{p})\right] .
\end{align}
The slow degrees of freedom are identified as $\mb{z}_s = (\mb{X},\mb{p})$ and the 
fast degrees of freedom as $\mb{z}_f = \mb{\tilde{v}}$.
We use the notation $\mathcal{A}_{\mathbf{\tilde v}}  \left[ \cdot \right]$ to represent contributions from the $\frac{d \mathbf{\tilde v}}{dt}$ equation to $L_{fast}$ since the term includes contributions of both first and second order in $\nabla_{\mathbf{ \tilde v}}$.

We split further the fast operator 
\begin{eqnarray}
L_{fast} & = & \frac{1}{\epsilon}\left(L_2 + \epsilon\tilde{L}_2\right) \\
\label{L_2}
L_{2} & = & [- \OpFSDissp C_1 \mathbf{\tilde{v}}] \cdot \nabla_{\mathbf{\tilde{v}}} + \kappa \OpFSDissp : \nabla_{\mathbf{\tilde{v}}}^2 \\
\tilde{L}_2 & = & - \alpha \kappa \nabla_{\mathbf{X}} \Phi(\mathbf{X})\cdot \nabla_{\mathbf{\tilde{v}}}  - \mathcal{A}_{\mathbf{\tilde v}}  \left[\nabla_{\mathbf{X},\mathbf{p}} \mathbf{v_0} \cdot \frac{d}{dt}(\mathbf{X},\mathbf{p})\right].
\end{eqnarray}
The Einstein summation convention for repeated indices is used throughout.  We denote by $A:B = A_{ij} B_{ij}$.  
The 
$\tilde{L}_2$ will be expressed more explicitly later.  For now, we remark that 
because of the stochastic contribution of the term ${d \mathbf{p}}/{dt}$, the differential 
operator $\tilde{L}_2$ when fully expressed is second order in $\mathbf{\tilde{v}}$.
     
In the perturbation analysis, it is sufficient to know $L_2$ to determine the stationary probability distribution
satisfying equation~\ref{equ_Psi_steady} for the fast degrees of freedom
\begin{align}
\label{psi_v}
\Psi(\mb{\tilde{v}}) = \f{\sqrt{\mbox{det}C}}{ (2 \pi)^{N/2}} \exp\left\lbrack
{-\f{1}{2 }\mathbf{\tilde{v}}^T C \mathbf{\tilde{v}} }\right\rbrack, \mbox{\hspace{0.8cm}} 
\text{ where }  
C   = \kappa^{-1} C_1.
\end{align}
This gives a Gaussian distribution for $\tilde{\mathbf{v}}$ with mean $\mathbf{0}$ and covariance $C^{-1}$.
The $N$ denotes the number of degrees of freedom for a configuration of the microstructures.  This solution intuitively corresponds to the Gibbs-Boltzmann distribution of $\mb{z}_f$ when holding 
the $\mb{z}_s$ degrees of freedom fixed (see Section~\ref{sec_general_reduction_procedure}). 

We split further the slow operator by
\begin{eqnarray}
L_{slow} & = & \bar{L}_1 + L_1  \\
\label{bar_L1}
\bar{L}_1 &= & \left(  \frac{1}{\delta}\OpFldDissp (\mathbf{p} - \kappa^{-1} \OpSFCouple [\mathbf{v}_0]) 
- \alpha\OpSFCouple \nabla_\mathbf{X} \Eng (\mathbf{X}) 
+ \kappa^{-1} (\nabla_{\mathbf{X}} \OpSFCouple [\mathbf{v}_0])\cdot \mathbf{v}_0 
+ \nabla_{\mathbf{X}} \OpSFCouple : C_1^{-1}  \right) \cdot \nabla_{\mathbf{p}} \mbox{\hspace{0.5cm}}\\
\nonumber
&-  & \frac{1}{\delta} \OpFldDissp: \nabla_{\mathbf{p}}^2 +  {\mathbf{v}_0} \cdot \nabla_{\mathbf{X}}  \\
\label{equ_L1_slow_ss}
L_1 & = & L_{slow} - \bar{L}_1. 
\end{eqnarray}
The $\bar{L}_1$ is obtained by averaging $L_{slow}$ with respect to $\Psi$.
We have used that the average of $\mathbf{v} = \mathbf{v}_0+\mathbf{\tilde{v}}$ with respect to $\Psi(\tilde{\mathbf{v}})$ is $\mathbf{v}_0$, which can be seen since $\mathbf{v}_0$ is a function of only slow degrees of freedom and the expression~\ref{psi_v} obtained for $\Psi$ is a Gaussian centered at zero.
Similarly, we  remark the $ \nabla_{\mathbf{X}} \OpSFCouple : C_1^{-1}$ term comes from the averaging of the term quadratic in $\mathbf{v}$.
  This splits the operator into a part $L_1$ that averages 
to zero  and a part $\bar{L}_1$ that may have a non-zero average.  The presented splittings into ``slow'' and ``fast'' parts
provide the required decomposition of the infinitesimal generators for our perturbation analysis.

We remark that the $\bar{L}_1$ operator describes the leading order dynamics for strong-coupling case considered. That is, the no-slip dynamics may already be recovered from $\bar{L}_1$ without determining the next-order terms. The next-order dynamics will be captured by the $\bar{L}_0$ term given by~\ref{equ_L_0}.

\subsubsection{Inverting the $L_2$ Operator}
To obtain the reduced stochastic process including the $\epsilon$-order term, we must determine the operator
\begin{eqnarray}
L_0 = -\int \Psi (L_1+ \tilde{L}_2) L_2^{-1} L_1 d\mb{z}_f. 
\end{eqnarray}
An often challenging step in determining $L_0$ is to perform the inverse of $L_2$ 
to find $w = -L_2^{-1} L_1 u_0$.  While in simple 
cases the resulting equation $L_2 w = -L_1 u_0$ can be solved directly, 
we take a more general approach by representing the action of the inverse 
operator over an orthonormal basis determined from solving a related 
Sturm-Liouville problem~\cite{Strauss2008}, see Appendix~\ref{sec_sl}.
To apply this approach we use that $L_2$ has the form
\begin{eqnarray}
\nonumber
L_2 & = & - (\OpFSDissp C \mathbf{\tilde{v}})_i \f{\p }{\p \tilde{v}_i} + \OpFSDissp_{ij}\f{\p^2}{\p \tilde{v}_i \p \tilde{v}_j}.
\end{eqnarray}
This can be put into an even more convenient form by choosing a change of basis for the velocity vector $\mathbf{\tilde{v}}$ so that the matrices diagonalize and the
operator is the sum $L_2 = \sum_i L^{(i)}_2$ where $L^{(i)}_2$ only involves independently the $i^{th}$ coordinate of velocity.  
For this purpose, we introduce the change of variable $\boldsymbol\alpha  = C^{{1}/{2}} \mathbf{\tilde{v}}$, where the 
square root $C^{{1}/{2}}$ is ensured
to exist since C is symmetric and positive semi-definite.  This allows us to express the operator as
\begin{eqnarray}
L_2 = A_{nm} \left(-\alpha_m \partial_{\alpha_n} + \partial_{\alpha_m}\partial_{\alpha_n}  \right), \mbox{\hspace{1cm}} 
A = C^{1/2} \Upsilon C^{1/2}.
\end{eqnarray}
Since $A$ is symmetric there is a unitary operator $Q$ for a a change of basis that diagonalizes the operator to
yield $D = Q^T A Q$ with $D_{ij} = \delta_{ij} d_i$, 
$\boldsymbol\beta = Q^T \boldsymbol\alpha$.  This gives 
\begin{eqnarray}
L_2 = \sum L_2^{(i)}, \mbox{\hspace{1cm}} 
L_2^{(i)} = d_i \left(-\beta_i \partial_{\beta_i} + \partial_{\beta_i}^2\right).
\end{eqnarray}
We remark that the cumulative change of variable used is $\boldsymbol\beta = Q^T C^{1/2} \mathbf{\tilde{v}}$.  We find it 
convenient also to introduce $\hat{Q} = C^{-1/2} Q $.
We can now express the inverse problem in terms of Sturm-Liouville operators, see Appendix~\ref{sec_sl}.  
This is achieved by introducing a factor to define a new operator
\begin{eqnarray}
\hat{L}_2 & = & e^{-\frac{1}{2}\boldsymbol{\beta}^2} L_2.
\end{eqnarray}
The inverse problem that needs to be solved becomes
\begin{eqnarray}
\hat{L}_2 w & = & \tilde{f}(\beta) \mu(\beta) 
\end{eqnarray}
where $\mu(\beta) = e^{-\frac{1}{2}\boldsymbol{\beta}^2}$ and $\tilde{f}  =  -L_1 u_0$.
This gives the eigenvalue problem
\begin{eqnarray}
\label{equ_eigenvalue_prob_beta}
\sum_i d_i\mu(\beta) \left(-\beta_i \partial_{\beta_i} + \partial_{\beta_i}^2\right) \phi_\mb{k}(\boldsymbol\beta) & = & \lambda_\mb{k}\mu(\beta) \phi_\mb{k}(\boldsymbol\beta).
\end{eqnarray}
The separated form of the differential operator allows for the solution to be represented in the separated form 
$\phi_{\mb{k}}(\boldsymbol{\beta}) = \prod_i \phi_{k_i}(\beta_i)$ with 
$k_i = \lbrack\mb{k}\rbrack_i$, $\beta_i = \lbrack\boldsymbol{\beta}\rbrack_i$.
The equation~\ref{equ_eigenvalue_prob_beta} can be decomposed into distinct 
Sturm-Liouville problems for the eigenfunctions $\phi_{k_i}(\beta_i)$ by using 
$p_{i}(\beta_i) = d_{i}e^{-\frac{1}{2}\boldsymbol{\beta_i}^2}$, see Appendix~\ref{sec_sl}.  
After some algebra, each of these eigenvalue problems have the general form 
\begin{eqnarray}
\label{equ_SL_hermite}
\phi_{k}''(\beta) -\beta \phi_{k}'(\beta) & = & 
\tilde{\lambda}_k \phi_{k}(\beta).
\end{eqnarray}
The $\beta$ is now simply a scalar variable.  In this case, we have the 
well-known Sturm-Liouville equations for the Hermite Orthogonal Polynomials~\cite{Strauss2008}.  
The eigenvalues can be shown to be the non-negative integers.  We denote the 
$k^{th}$ Hermite Polynomial by $H_k(\beta)$ and the eigenvalue by 
$\tilde{\lambda}_k = k \geq 0$ with $k\in\mathbb{Z}^{+}$.   
For equation~\ref{equ_eigenvalue_prob_beta} this gives the eigenfunctions 
$\phi_\mb{k}(\boldsymbol\beta) = \prod_i H_{k_i}(\beta_i)$ and the eigenvalues 
$\lambda_\mb{k} = \sum_i d_i \tilde{\lambda}_{k_i} = \sum_i d_i {k_i}$.  The 
action of the inverse operator on a basis element can then be expressed as 
\begin{eqnarray}
L_2^{-1} \phi_\mb{k} = -\left\lbrack \sum_i d_i {k_i} \right\rbrack^{-1} \phi_\mb{k}.
\end{eqnarray}
In the case that $\tilde{f} = -L_1 u_0$ is a polynomial of finite degree in $\mb{z}_f$, the 
inverse is also a finite degree polynomial.  For the low degree polynomials that arise 
from the SELM dynamics these results provide a particularly useful inversion procedure.  
For convenience, we list here the first few Hermite Polynomials
\begin{eqnarray}
H_0(\beta) = 1, \mbox{\hspace{1cm}} 
H_1(\beta) = \beta, \mbox{\hspace{1cm}} 
H_2(\beta) = \beta^2 - 1. 
\end{eqnarray}
A few useful inversion formulas of which we shall make use include
\begin{eqnarray}
\label{equ_inv_one}
L_2^{-1} \beta_i  & = & -D_{ij}^{-1} \beta_j = -d_{i}^{-1} \beta_i \\
\label{equ_inv_inotj}
L_2^{-1} \beta_i\beta_j & = & -\left(d_{i} + d_{j}\right)^{-1} \beta_i\beta_j, \hspace{0.5cm} i \not= j \\
\label{equ_inv_iisj}
L_2^{-1} \left( \beta_i^2 - 1 \right) & = & -(2d_{i})^{-1} \left( \beta_i^2 - 1 \right). 
\end{eqnarray}
In some of the calculations it is helpful to use the tensor notation $D_{ij}^{-1} = d_i^{-1} \delta_{ij}$
and to combine equation~\ref{equ_inv_inotj} and~\ref{equ_inv_iisj} to obtain 
\begin{eqnarray}
\label{equ_inv_comb}
L_2^{-1} \left(\beta_i\beta_j - \delta_{ij}\right) & = & 
-\left(d_{i} + d_{j}\right)^{-1} \left(\beta_i\beta_j - \delta_{ij}\right) 
= -E_{ij}\left(\beta_i\beta_j - \delta_{ij}\right)
\end{eqnarray}
where $E_{ij} = \left(d_{i} + d_{j}\right)^{-1}$.

\subsubsection{Representation of the Operators under the Change of Variable for Strong Coupling}
To succinctly carry out the calculation of the effective infinitesimal generator, it is helpful to 
introduce some notation for the change of variable we use from $\tilde{\mb{v}}$ to $\boldsymbol{\beta}$.  
To summarize the notation we introduced so far we had
$A \equiv C^{1/2} \Upsilon C^{1/2}$,
$D = Q^T A Q \text{ (Q unitarily diagonalizes A),}$
$\hat{Q} \equiv  C^{-1/2} Q$,  $\text{ so that } D ^{-1}= \hat{Q}^T \Upsilon^{-1} \hat{Q}$, 
$ \hat{Q} \boldsymbol{\beta}  \equiv\mathbf{\tilde{v}}.$
To account for the drift contributions to the slow variable in a form amenable to Hermite polynomials of order $0$, $1$, and $2$ 
we introduce respectively
\begin{align}
\label{def_T}
T & \equiv  \mathbf{v}_0 \boxtimes \bigg[{\frac{1}{\delta}\OpFldDissp} \left(\mathbf{{p}} - \kappa^{-1}  {\OpSFCouple}[\mathbf{v}_0]\right) -\alpha {\OpSFCouple}[\nabla_{{X}}{\Phi}(\mathbf{{X}})] 
                         + \nabla_{\mathbf{X}} \OpSFCouple : C_1^{-1} 
			+\kappa^{-1}  (\nabla_{\mathbf{X}} \OpSFCouple [\mathbf{v}_0])\cdot \mathbf{v}_0 \bigg] \\
R & \equiv \mathcal{I}_{N\times N} \boxtimes B \text{,\hspace{0.25cm} with} 
\label{def_B} \mbox{\hspace{0.2cm}} B \equiv  \kappa^{-1} \bigg[- \frac{1}{\delta} \OpFldDissp \OpSFCouple (\cdot)
		+ (\nabla_{\mathbf{X}} \OpSFCouple(\cdot) ) \cdot \mathbf{v}_0 
		+ \nabla_{\mathbf{X}} (\OpSFCouple \mathbf{v}_0) \cdot (\cdot) \bigg]\\
\label{def_S} 
S & \equiv  0_{N\times N \times N} \boxtimes \kappa^{-1}  \nabla_{\mathbf{X}} \OpSFCouple.
\end{align}
The notation $\boxtimes$ is introduced to ``glue-together'' two tensors along the first index, so that for $A_{ijk}$ with $1 \leq i \leq N$ and $B_{ijk}$ with $1 \leq i \leq M$ we define the new tensor $C = A \boxtimes B$ by $C_{ijk} = A_{ijk}$ when $1 \leq i \leq N$ and $C_{ijk} = B_{{(i-N)}jk}$ when $N + 1 \leq i \leq N + M$.  We call $\boxtimes$ the ``glue-operator.'' 
The $0_{N\times N \times N}$ denotes a 3-tensor of zeros and 
$\mathcal{I}_{N\times N}$ the identity 2-tensor.  The order of coordinates in~\ref{def_S} is understood to be $(\nabla_{\mathbf{X}} \OpSFCouple)_{ijk} = \partial_{X_{k}} \OpSFCouple_{ij}$. It will be useful in $\tilde{L}_2$ to introduce the 
modified noise term
\begin{eqnarray}
V &
= 0_{N\times N} \boxtimes - C_1^{-1}   \OpFSCouple  \OpFldDissp \OpSFCouple C_1^{-1}.
\label{def_V}
\end{eqnarray}
where $0_{N \times N}$ denotes a 2-tensor of zeros.
Here, $C_1$ is the same as in~\ref{equ_C_1}.
We also use 
\begin{eqnarray}
\mathbf{y} \equiv \mathbf{z_s} = (\mathbf{X},\mathbf{p}).
\end{eqnarray}
This allows for the slow operator of equation~\ref{equ_L1_slow_ss} to be expressed succinctly as 
\begin{eqnarray}
L_1 &= & 
\left[ R_{ij} \tilde{v}_j +S_{ijk} (\tilde{v}_k \tilde{v}_j - \overline{\tilde{v}_k \tilde{v}_j}) \right] \f{\p}{\p y_i}.
\label{L1}
\end{eqnarray}
Ultimately, this operator will be expressed in terms of a change of variable from $\mathbf{\tilde{v}}$ to $\boldsymbol{\vOrth}$.
This makes it useful to make also the change of variable for $R$ and $S$, which is given by 
\begin{align}
\label{def_R_hat}
\hat{R}_{ij} = & R_{ik} \hat{Q}_{kj} &
\hat{S}_{ijk} =& S_{ilm} \hat{Q}_{lj} \hat{Q}_{mk}. 
\end{align}
We then have for the fast operator an expression in terms of $\boldsymbol{\vOrth}$
\begin{eqnarray}
\label{equ_def_L1_i}
L_1  &= & \left[ \hat{R}_{ij} \vOrth_j +\hat{S}_{ijk} (\vOrth_k \vOrth_j - \delta_{kj} ) \right] \f{\p }{\p y_i} = L_1^{(1)} + L_1^{(2)} \hspace{0.25cm} \text{where }\\
L_1^{(1)} & =& \hat{R}_{ij} \vOrth_j \f{\p }{\p y_i}, \hspace{1.25cm}
L_1^{(2)} = \hat{S}_{ijk} (\vOrth_k \vOrth_j - \delta_{kj} ) \f{\p }{\p y_i}.
\end{eqnarray}
In the interest of computing the $\mathcal{A}_{\mathbf{\tilde v}}  [\nabla_\mathbf{y} \mathbf{v}_0 \cdot \frac{d\mathbf{y}}{dt}]$
 appearing in $\tilde{L}_2$, it is useful to express 
\begin{eqnarray}
\frac{d\mathbf{y} }{dt} 
& = & U(\mathbf{\tilde{v}}) + \mathbf{g}_\text{thm} \text{,\hspace{0.25cm} with}\\
\left[U(\mathbf{\tilde{v}})\right]_i & = &   T_i + R_{ij} \tilde{v}_j +S_{ijk} (\tilde{v}_k \tilde{v}_j - \overline{\tilde{v}_k \tilde{v}_j}) \\
\langle \mathbf{g}_{\mbox{\tiny thm}}(s), \mathbf{g}_{\mbox{\tiny thm}}^T (t)\rangle & = & 0_{N\times N} \boxtimes -2 \OpFldDissp \delta(t-s).
\end{eqnarray}
This allows for the fast operator to be expressed in terms of $\boldsymbol{\vOrth}$ as
\begin{eqnarray}
\tilde{L}_2
& = & 
\bigg[- \alpha\kappa \nabla_{\mathbf{X}} \Phi(\mathbf{X})
		 - \nabla_{\mathbf{y}} \mathbf{v_0}\cdot U(\mathbf{\tilde{v}}) \bigg] \cdot \nabla_{\mathbf{\tilde{v}}} 
		- \frac{1}{\delta} V : \nabla^2_{\mathbf{\tilde{v}} } \\
 & = &  \bigg[-\alpha \kappa \nabla_{\mathbf{X}} \Phi(\mathbf{X}) 
		- \nabla_{\mathbf{y}} \mathbf{v_0}\cdot U(\mathbf{\tilde{v}}) \bigg] \cdot \hat{Q} ^ {-T} \nabla_{\boldsymbol{\vOrth}} 
		- \frac{1}{\delta}\hat{Q}^{-1} V  \hat{Q}^{-T}: \nabla^2_{\boldsymbol{\vOrth}} \\
 & = &  \bigg[ \tilde{T}_i + \tilde{R}_{ij} \vOrth_j +\tilde{S}_{ijk} (\vOrth_k \vOrth_j - \delta_{kj}) \bigg] \cdot \nabla_{\boldsymbol{\vOrth}} 
		+ \tilde{V}: \nabla^2_{\boldsymbol{\vOrth}}. 
\end{eqnarray}
In these expressions we define
\begin{align}
\tilde{T} &= \hat{Q}^{-1}  (- \alpha\kappa \nabla_{\mathbf{X}} \Phi(\mathbf{X}) 
		- \nabla_{\mathbf{y}} \mathbf{v_0}\cdot T),  \mbox{\hspace{0.1cm}} 
\label{def_R_tilde}
&\tilde{R}_{mn} &= \hat{Q}^{-1}_{mi}   (-\partial_{y_j} (v_0)_i) \hat {R}_{jn},   \\
\label{def_S_tilde}
\tilde{S}_{lmn} &= \hat{Q}^{-1}_{li}  (-\partial_{y_j} (v_0)_i)\hat S_{jmn},   \mbox{\hspace{0.1cm}} 
&\tilde{V} &= -\frac{1}{\delta} \hat{Q}^{-1} V \hat{Q}^{-T}. 
\end{align}
The $V$ term in $\tilde L_2$ which we defined in~\ref{def_V} results from the form of $\mathbf{g}_\text{thm}$ and the relationship~\ref{equ_v0}.

Finally, we split $\tilde{L}_2$ for convenience in later integral expressions into several components that each involve a different degree polynomial in $\boldsymbol{\beta}$.  We label these using the convention that a derivative in $\boldsymbol{\beta}$ contributes ``negatively'' to the degree while multiplication by a variable in $\boldsymbol{\beta}$ contributes ``positively'' to the degree.  This gives the decomposition
\begin{eqnarray}
\tilde{L}_2 = \sum \tilde{L}_2^{(i)}, \mbox{\hspace{3cm}}   \\
\label{equ_def_L2_i}
\tilde{L}_2^{(-2)} =  \tilde{V}_{ij} \p^2_{\vOrth_{ij}}, \mbox{\hspace{0.3cm}}
\tilde{L}_2^{(-1)} =  \tilde{T}_i \p_{\beta_i}, \nonumber \mbox{\hspace{2cm}}\\
\tilde{L}_2^{(0)} =  \tilde{R}_{ij} \vOrth_j \p_{\beta_i}, \mbox{\hspace{0.3cm}} 
\tilde{L}_2^{(1)} =  \tilde{S}_{ijk} (\vOrth_k \vOrth_j - \delta_{kj})\p_{\beta_i}. \mbox{\hspace{0cm}}
\end{eqnarray}
These conventions provide useful notation to succinctly express the consequences of the 
change of variable from $\tilde{\mb{v}}$ to $\boldsymbol{\vOrth}$.

\subsubsection{Computing the Effective Infinitesimal Generator for Strong Coupling} 
To obtain the effective infinitesimal generator $\bar{L} = \bar{L}_1 + \epsilon L_0$ we must 
still compute $L_0 = -\int \Psi (L_1 + \tilde{L}_2) L_2^{-1} L_1 d\mb{z}$.
We assume in this section that $\Gamma \Lambda$ does not depend on $\mathbf{X}$ to leading order.  This corresponds to the mobility being approximately constant and has the consequence that $C$ is approximately constant in $\mathbf{X}$.
We start by expressing the probability distribution $\Psi$ from equation~\ref{psi_v} in terms of the variable 
$\boldsymbol\beta$ and use the associated Jacobian to obtain
\begin{equation}
\Psi(\boldsymbol\beta) = {{(2 \pi)}^{-N/2}} \exp\left\lbrack{-\frac{1}{2} \boldsymbol\beta^2}\right\rbrack.
\label{psi_new_var_strongClimit}
\end{equation}
To determine the operator is useful to split into the parts $L_0 = \sum_{ij} I_{ij} + \sum_{ij} J_{ij}$ with
\begin{align}
I_{ij} = - \int \Psi(\boldsymbol\beta) L_1^{(i)} L_2^{-1} L_1^{(j)} d\boldsymbol\beta, \mbox{\hspace{1cm}}
J _{ij}= -  \int \Psi(\boldsymbol\beta) \tilde{L}_2^{(i)} L_2^{-1} L_1^{(j)} d\boldsymbol\beta.
\end{align}
The operators $L_1^{(i)}$ and $\tilde{L}_2^{(i)}$ are defined in equations~\ref{equ_def_L1_i} and~\ref{equ_def_L2_i}.
In practice, the terms $I_{11}$ and $I_{22}$ are the only $I_{ij}$ needed to determine $L_0$ since 
the $I_{ij} = 0$ when $i \not= j$.  This is a consequence of odd degree monomials in $\beta$ averaging to 
zero under the probability distribution.  Similarly, the only terms $J_{ij}$
that are non-zero and needed to determine $L_0$ are $J_{-2,2},J_{0,2}, J_{-1,1},J_{1,1}$.

A useful feature of our decomposition is that the operators involve terms that are at most a degree two multinomial 
in the variables $\beta_i$.  From the inversion formulas established in equations~\ref{equ_inv_one}--~\ref{equ_inv_iisj}
and the decomposition of $L_1$ we have
\begin{align}
I_{11} & =  -  \int_{\mathbb{R}^N}  \Psi(\mathbf{\vOrth}) \hat{R}_{nm} \vOrth_m \p_{y_n} L_2^{-1} [\hat{R}_{ij} \vOrth_j \p_{y_i}] d\boldsymbol\beta \\
&=   \int_{\mathbb{R}^N}  \Psi(\mathbf{\vOrth}) \hat{R}_{nm} \vOrth_m \p_{y_n} [\hat{R}_{ij} D_{jk}^{-1} \vOrth_k \p_{y_i}] d \boldsymbol\beta
= \hat{R}_{nk} D_{jk}^{-1} \p_{y_n}[\hat{R}_{ij} \p_{y_i}]. 
\end{align}
We used the specific inversion formula~\ref{equ_inv_one} to obtain that 
$L_2^{-1} [\hat{R}_{ij} \vOrth_j \p_{y_i}] = [\hat{R}_{ij} D_{jk}^{-1} \vOrth_k \p_{y_i}]$.
We can reverse the change of variable to express this in terms of the original tensors associated with the variables $(\mb{X},\mb{p})$ as
\begin{align}
\label{equ_comp_I_11}
I_{11} &=( R C^{-1/2} Q D^{-1} Q^T C^{-1/2} )_{nj} \p_{y_n} [R_{ij} \p_{y_i} ] \\
 &= (R C^{-1} \OpFSDissp^{-1} C^{-1})_{nj} [\p_{y_n} R_{ij}\p_{y_i}  + R_{ij} \p^2_{y_n y_i}] \\
 &= (R C^{-1} \OpFSDissp^{-1} C^{-1})_{nj} [\p_{y_n} B_{ij} \p_{p_i} ] 
+  (R C^{-1} \OpFSDissp^{-1} C^{-1} R^T)_{ni} [ \p^2_{y_n y_i}] \\
&=r \cdot \nabla_{\mathbf{p}}
+ (q q^T) : \nabla^2_{\mathbf{y}}.
\end{align}
with 
\begin{align}
\label{equ_r }
r = (\nabla_{\mathbf{X},\mathbf{p}} B): (C^{-1} \OpFSDissp^{-1} C^{-1} R^T), &&
 q = R C^{-1} \OpFSDissp^{-1/2}.
\end{align}
  We also used that
the first $N$ rows of $R$ are constant (and correspond to $\mathcal{I}$).  The $B$ is defined in equation~\ref{def_B}.   

We next compute $I_{22}$.  From equation~\ref{def_S}, we denote 
$\hat{S}_{ijk} =S_{imn} \hat{Q}_{mj} \hat{Q}_{nk}$.  To avoid confusion in the notation for the indices 
$m$ and $n$, we denote this sum explicitly.  This gives
\begin{align}
I_{22} &=  - \int_{\mathbb{R}^N} \Psi(\mathbf{\vOrth}) L_1^{(2)} L_2^{-1} L_1^{(2)} d \boldsymbol\beta\\
&=  -  \int_{\mathbb{R}^N}  \Psi(\mathbf{\vOrth}) \hat{S}_{ijk} (\vOrth_j \vOrth_k - \delta_{jk} ) \p_{y_i} L_2^{-1} [\hat{S}_{lmn} (\vOrth_m \vOrth_n - \delta_{mn} ) \p_{y_l}] d \boldsymbol\beta \\
&=  \int_{\mathbb{R}^N}  \Psi(\mathbf{\vOrth}) \hat{S}_{ijk} (\vOrth_j \vOrth_k - \delta_{jk} ) \p_{y_i} \left[\sum_{mn} \hat{S}_{lmn} E_{mn} (\vOrth_m \vOrth_n - \delta_{mn} ) \p_{y_l} \right] d \boldsymbol\beta \\
&=\sum_{mn} \left[E_{mn}
 \int_{
\mathbb{R}^N}  \Psi(\mathbf{\vOrth}) (\vOrth_j \vOrth_k - \delta_{jk} )  \left[ (\vOrth_m \vOrth_n - \delta_{mn} ) \right] d \boldsymbol\beta
\left( \hat{S}_{ijk}^{(2)} \hat{S}_{lmn}^{(2)}   \p^2_{p_i p_l}  + 
\hat{S}_{ijk}^{(2)}  \partial_{X_i} \hat{S}_{lmn}^{(2)} \partial_{p_l}
\right]
\right).
\end{align}
The inversion formula~\ref{equ_inv_comb} was used to obtain 
$L_2^{-1} (\vOrth_m \vOrth_n - \delta_{mn}) = E_{mn} (\vOrth_m \vOrth_n - \delta_{mn})$.
Another important point to mention is that $\hat{S}_{ijk}$ only yields non-zero terms when $i>N$, and that $\hat{S}_{ijk}$ is a function of $\mathbf{X}$ but not $\mathbf{p}$.  This follows since the indices with $i < N$ involve contributions to the $\mb{X}$ equations which are zero and were represented using our glue-operator in equation~\ref{def_S}. For this reason it is convenient to use the notation for $\hat{S}$ above, $A_{ijk}^{(2)} = A_{(i-N) jk}$ for $i>N$.

To integrate the expressions against $\Psi$, we find it useful to introduce an integration by parts in the variable $\vOrth_j$
\begin{align}
&\int_{\mathbb{R}^N}  \Psi(\mathbf{\vOrth}) \vOrth_j \vOrth_k  \vOrth_m \vOrth_n  d\boldsymbol\beta \\
&= \frac{1}{{(2 \pi)}^{N/2}} \int_{\mathbb{R}^{N-1}} 
\left( \int_{\mathbb{R}} \vOrth_j e^{-\frac{1}{2} \vOrth_j^2}  \vOrth_k  \vOrth_m \vOrth_n d \vOrth_j \right) e^{-\frac{1}{2} \sum_{a \neq j}\vOrth_a^2 } d \boldsymbol\beta^{/j} \\
&= \frac{1}{{(2 \pi)}^{N/2}} \int_{\mathbb{R}^N}  e^{-\frac{1}{2} \vOrth_j^2}  \p_{\vOrth_j}(\vOrth_k  \vOrth_m \vOrth_n) d \vOrth_j e^{-\frac{1}{2} \sum_{a \neq j}\vOrth_a^2 } d \boldsymbol\beta^{/j} \\
&= \delta_{jk} \delta_{mn} + \delta_{jm} \delta_{kn} + \delta_{jn} \delta_{km}.
\end{align}
The $d \boldsymbol\beta^{/j} = d\vOrth_1 \cdots d\vOrth_{j - 1}d\vOrth_{j + 1} \cdots d\vOrth_{N}$ denotes the differential 
excluding $dw_j$.  Using this result we obtain  
\begin{align}
&\int_{\mathbb{R}^N}  \Psi(\mathbf{\vOrth}) (\vOrth_j \vOrth_k - \delta_{jk})  (\vOrth_m \vOrth_n - \delta_{mn})  d \boldsymbol\beta\\
&= \delta_{jk} \delta_{mn} + \delta_{jm} \delta_{kn} + \delta_{jn} \delta_{km} - 2 \delta_{jk}\delta_{mn} + \delta_{jk}\delta_{mn} \\
& = \delta_{jm} \delta_{kn} + \delta_{jn} \delta_{km}.
\end{align}
This yields
\begin{align}
\label{I_2_1_new}
I_{22} &= \sum_{mn}\left[
\left(
 \hat{S}_{ijk}^{(2)} \hat{S}_{lmn}^{(2)}\p^2_{p_i p_l} 
+ \hat{S}_{ijk}^{(2)}  \partial_{X_i} \hat{S}_{lmn}^{(2)} \partial_{p_l} \right)
\left( E_{mn} (\delta_{jm} \delta_{kn} + \delta_{jn} \delta_{km})  \right)
\right]
\\
\label{I_2_2_new} 
& = \sum_{mn} \left\{
E_{mn} \left[
\left(
\hat{S}_{imn}^{(2)} \hat{S}_{lmn}^{(2)} + \hat{S}_{inm}^{(2)} \hat{S}_{lmn}^{(2)}
\right)  \p^2_{p_i p_l}
+
\left(
\hat{S}_{imn}^{(2)} 
+ \hat{S}_{inm}^{(2)} 
\right) \partial_{X_i} \hat{S}^{(2)}_{lmn}
 \partial_{p_l}
\right]
\right\}.	
\end{align}
Denote the first- and second-order components of $I_{22}$ by $I_{22}^{(1)}$ and $I_{22}^{(2)}$, respectively.
Since $E_{mn}$ is symmetric, we can write the the second-order  differential operator as
\begin{align}
\label{I_2_1_new2}
I_{22}^{(2)} & = \sum_{mn}  E_{mn}
\left(
\hat{S}^{(2)}_{imn}\hat{S}^{(2)}_{lmn}
+
 \frac{1}{2} \hat{S}_{imn}^{(2)} \hat{S}_{lnm}^{(2)} + \frac{1}{2} \hat{S}_{lmn}^{(2)}\hat{S}_{inm}^{(2)} 
\right)
 \p^2_{p_i p_l}.
\end{align}
This gives
 $\frac{1}{2}A_{il}\p^2_{p_i p_l}$ with a tensor $A$ that is symmetric in the indices $i,l$.  This allows for the 
operator to be expressed as
\begin{equation}
\label{I_2_5_new}
I_{22}^{(2)} = \frac{1}{2} [\sigma \sigma^T]_{il} \p^2_{p_i p_l}  = \frac{1}{2} \sigma \sigma^T : \nabla_{\mathbf{p}}^2
\end{equation}
where $\sigma$ is a factor such that $A = \sigma \sigma^T$.

To determine the specific form of $\sigma$, we consider for fixed indices $m$ and $n$
\begin{align}
\label{P_remove_hat_new}
\mathcal{W}_{ilmn} = \hat{S}_{imn}^{(2)} \hat{S}_{lmn}^{(2)} + \hat{S}_{imn}^{(2)} \hat{S}_{lnm}^{(2)} = S_{iab}^{(2)} S_{lcd}^{(2)} \hat{Q}_{am} \hat{Q}_{bn} (\hat{Q}_{cm} \hat{Q}_{dn} + \hat{Q}_{cn} \hat{Q}_{dm}).
\end{align}
Using the form \ref{P_remove_hat_new} in \ref{I_2_2_new}, we have
\begin{align}
\label{I_2_3_new}
I_{22}^{(2)}& =  
S_{iab}^{(2)} S_{lcd}^{(2)}  \left[\sum_{mn}\hat{Q}_{am} \hat{Q}_{bn} E_{mn} (\hat{Q}_{cm} \hat{Q}_{dn} + \hat{Q}_{cn} \hat{Q}_{dm}) \right]   
\p^2_{p_i p_l}.
\end{align}
By interchanging $m$ and $n$ and averaging the original and new forms of $I_{22}$ we find
\begin{align}
\label{equ_comp_I_22}
I_{22}^{(2)} & = \frac{1}{2} S_{iab}^{(2)} S_{lcd}^{(2)}  \left[\sum_{mn}(\hat{Q}_{am} \hat{Q}_{bn} + \hat{Q}_{an} \hat{Q}_{bm}) E_{mn} (\hat{Q}_{cm} \hat{Q}_{dn} + \hat{Q}_{cn} \hat{Q}_{dm}) \right]   \p^2_{p_i p_l}.
\end{align}
This gives 
\begin{align}
\label{equ_sigma}
\sigma_{i,\{m,n\}} = S_{iab}^{(2)} (\hat{Q}_{am} \hat{Q}_{bn} + \hat{Q}_{an} \hat{Q}_{bm}) \sqrt{E_{mn}}.
\end{align}
We remark that the summation convention is assumed on the indices $a,b$, but not $m,n$.  This provides an explicit form for
the factor $A = \sigma\sigma^T$ required in equation~\ref{equ_sigma}.  
A similar computation gives
\begin{align}
\label{equ_comp_I_22_2}
I_{22}^{(1)} & = \frac{1}{2} S_{iab}^{(2)} \partial_{X_i} S_{lcd}^{(2)}  \left[\sum_{mn}(\hat{Q}_{am} \hat{Q}_{bn} + \hat{Q}_{an} \hat{Q}_{bm}) E_{mn} (\hat{Q}_{cm} \hat{Q}_{dn} + \hat{Q}_{cn} \hat{Q}_{dm}) \right]   \p^2_{p_l} \\
&= \frac{1}{2} \sum_{i, \{m,n\}} \partial_{X_i} \sigma_{l, \{m,n\}} \sigma_{i, \{m,n\} } \partial _{p_l}^2  =\left[\frac{1}{2} \nabla_{\mathbf{X}} \sigma : \sigma^T\right]\cdot \nabla_{\mathbf{p}}
.
\end{align}
The double dot-product is understood to be taken over $i,\{m,n\}$ with our convention. Above we include the explicit sum over $i$ as well as $\{m,n\}$ to make this clear.
To summarize, we have
\begin{equation}
\label{equ_comp_I_22_tot}
I_{22} = \frac{1}{2} \sigma \sigma^T : \nabla_{\mathbf{p}}^2 +
\left[\frac{1}{2} \nabla_{\mathbf{X}} \sigma : \sigma^T\right]\cdot \nabla_{\mathbf{p}}
.
\end{equation}

This result is useful since it provides an 
explicit form in the reduced equations for any general choice that is made for the coupling operator 
$\Upsilon$.   

Next, we can compute the integrals $J_{ij}$.  Recalling the function on which the operator is applied depends only on the slow variables, we may apply the derivative in $\tilde{L}_2$ only on the fast variables appearing on $L_2^{-1} L_1$. We find
\begin{align}
\label{equ_comp_Jn2n2}
J_{-2,2} &=  - \int_{\mathbb{R}^N} \Psi(\mathbf{\vOrth}) \tilde{L}_2^{(-2)} L_2^{-1} L_1^{(2)} d\boldsymbol\beta\\
&=    \int_{\mathbb{R}^N}  \Psi(\mathbf{\vOrth}) \tilde{V}_{ij} \p^2_{\vOrth_{ij}} \left[\sum_{m,n} \hat{S}_{lmn} E_{mn} (\vOrth_m \vOrth_n - \delta_{mn} ) \p_{y_l} \right] d \boldsymbol\beta \\
&=   \int_{\mathbb{R}^N}  \Psi(\mathbf{\vOrth}) \tilde{V}_{ij}  \left[\sum_{mn} \hat{S}_{lmn} E_{mn}  (\delta_{jm} \delta_{ni}+ \delta_{mi} \delta_{jn}) \p_{y_l} \right] d \boldsymbol\beta \\
&=   \sum_{mn} \hat{S}_{lmn} E_{mn}  (\tilde{V}_{nm} + \tilde{V}_{mn}  )  \p_{y_l} \\
&= \frac{1}{2} \sum_{m,n} S_{lab} \left( \hat Q_{am} \hat Q_{bn} +  \hat Q_{bm} \hat Q_{an}\right) E_{mn} (\tilde V_{mn} + \tilde V_{nm} ) \p_{y_l}. \\
&= \sum_{m,n} S_{lab} \left( \hat Q_{am} \hat Q_{bn} +  \hat Q_{bm} \hat Q_{an}\right) E_{mn} \tilde V_{mn}   \p_{y_l}. \\
&= \sum_{m,n}  \sigma_{l,\{m,n\}} \nu_{mn} \p_{y_l} 
= Y \cdot \nabla_{\mathbf{p}}.
\end{align}
We label the tensors 
\begin{align}
\nu_{mn} = \sqrt{E_{mn}} \tilde V_{mn},  &&
Y_i  = \sum_{m,n} \sigma_{i,\{m,n\}} \nu_{mn}.
\label{equ_Y} 
\end{align}
Above, $\tilde V$ is given in~\ref{def_S_tilde} and $\sigma$ is given in ~\ref{equ_sigma}.
Next,
\begin{align}
\label{equ_comp_J02}
J_{0,2} & = - \int_{\mathbb{R}^N} \Psi(\mathbf{\vOrth}) \tilde{L}_2^{(0)} L_2^{-1} L_1^{(2)} d \boldsymbol\beta \\
&=    \int_{\mathbb{R}^N}  \Psi(\mathbf{\vOrth}) \tilde{R}_{ij} \vOrth_j \p_{\vOrth_{i}} \left[\sum_{m,n} \hat{S}_{lmn} E_{mn} (\vOrth_m \vOrth_n - \delta_{mn} ) \p_{y_l} \right] d \boldsymbol\beta \\
&=    \int_{\mathbb{R}^N}  \Psi(\mathbf{\vOrth}) \tilde{R}_{ij}  \left[\sum_{m,n} \hat{S}_{lmn} E_{mn} (\delta_{mi}\vOrth_j  \vOrth_n + \vOrth_m \vOrth_j  \delta_{ni}) \p_{y_l} \right] d \boldsymbol\beta\\
&=     \tilde{R}_{ij}  \left[\sum_{m,n} \hat{S}_{lmn} E_{mn} (\delta_{mi}\delta_{jn} + \delta_{mj} \delta_{ni}) \p_{y_l} \right] \\
&= \frac{1}{2} {S}_{lab}  \left[\sum_{m,n} (\hat{Q}_{am} \hat{Q}_{bn} + \hat{Q}_{an} \hat{Q}_{bm}) E_{mn}
 (\tilde{R}_{mn} + \tilde{R}_{nm})   \right]\p_{y_l} \\
&= {S}_{lab}  \left[\sum_{m,n} (\hat{Q}_{am} \hat{Q}_{bn} + \hat{Q}_{an} \hat{Q}_{bm}) E_{mn}
 \tilde{R}_{mn}   \right]\p_{y_l} \\
&= \sum_{m,n} \sigma_{l, \{m,n \} } \omega_{mn} \partial_{{p_l}} = Z \cdot \nabla_{\mathbf{p}}
\end{align}
We label the  tensors
\begin{align}
\omega_{mn} = \sqrt{E_{mn}} \tilde{R}_{mn}, && 
Z_i = \sum_{m,n} \sigma_{i, \{m,n \} } \omega_{mn}.
\label{equ_omega}
\end{align}
Here, $\tilde R$ is given in~\ref{def_R_tilde} and $\sigma$ is given in~\ref{equ_sigma}.
\begin{align}
\label{equ_comp_Jn11}
J_{-1,1}  =& - \int_{\mathbb{R}^N} \Psi(\mathbf{\vOrth}) \tilde{L}_2^{(-1)} L_2^{-1} L_1^{(1)} d \boldsymbol\beta
=   \int_{\mathbb{R}^N}  \Psi(\mathbf{\vOrth}) \tilde{T}_{i} \p_{\vOrth_{i}} [\hat{R}_{lj} D_{jk}^{-1} \vOrth_k \p_{y_l}] d \boldsymbol\beta\\
= & \int_{\mathbb{R}^N}  \Psi(\mathbf{\vOrth}) \tilde{T}_{i}  [\hat{R}_{lj} D_{ji}^{-1} \p_{y_l}] d \boldsymbol\beta
=  \hat{R}_{lj} D_{ji}^{-1} \tilde{T}_{i}   \p_{y_l} \\
=& - [R C^{-1} \Upsilon ^{-1}\gamma(\alpha \kappa)]\cdot \nabla_{\mathbf{y}}
= - [ C^{-1} \Upsilon ^{-1} \gamma(\alpha \kappa)]
	\cdot \nabla_{\mathbf{X}} 
 - [B C^{-1} \Upsilon ^{-1} \gamma(\alpha \kappa) ]\nonumber
	 \cdot \nabla_{\mathbf{p}} ,
\end{align}
where we have abbreviated 
\begin{align}
\label{def_gamma}
\gamma (\alpha \kappa) = ( \alpha \kappa \nabla_{\mathbf{X}} \Phi(\mathbf{X}) 
	+ \nabla_{\mathbf{y}} \mathbf{v_0}\cdot T).
\end{align}
Here, $T$ is given in~\ref{def_T}.
We emphasize the dependence on $\alpha \kappa$ for the discussion below on the relationship between the non-dimensional constants.
\begin{eqnarray}
\label{equ_comp_J11}
J_{1,1} & =& - \int_{\mathbb{R}^N} \Psi(\mathbf{\vOrth}) \tilde{L}_2^{(1)} L_2^{-1} L_1^{(1)} d \boldsymbol\beta \\
 &= & \int_{\mathbb{R}^N}  \Psi(\mathbf{\vOrth})  \tilde{S}_{inm} (\vOrth_n \vOrth_m - \delta_{nm})[\hat{R}_{lj} D_{ji}^{-1}  \p_{y_l}] d \boldsymbol\beta  = 0.
\end{eqnarray}
The last integral is $0$ since $\overline{\beta_n \beta_m} = \delta_{nm}$.

By combining the above results $L_0 = I_{11} + I_{22} + J_{-2,2} + J_{0,2} + J_{-1,1} + J_{1,1}$ from equations~\ref{equ_comp_I_11},~\ref{equ_comp_I_22_tot},~\ref{equ_comp_Jn2n2},~\ref{equ_comp_J02},~\ref{equ_comp_Jn11}, and~\ref{equ_comp_J11} we obtain the operator 
\begin{align}
&L_0 =\left[  r
	+ \frac{1}{2} \nabla_{\mathbf{X}} \sigma : \sigma^T
	+Y + Z - B C^{-1} \Upsilon ^{-1} \gamma(\alpha \kappa)
\right] \cdot \nabla_{\mathbf{p}} \\
	& - [ C^{-1} \Upsilon ^{-1} \gamma(\alpha \kappa)]
	\cdot \nabla_{\mathbf{X}}  \nonumber
	+ (q q^T) : \nabla^2_{\mathbf{y}}
	+\frac{1}{2} \sigma \sigma^T : \nabla_{\mathbf{p}}^2 
\hspace{1cm}
\end{align}

These results combined with the $\bar{L}_1$ given in equation~\ref{bar_L1} give the 
final reduced operator $L_0 = \bar{L}_1 + \epsilon L_0$.  While the $\bar{L}_1$ operator yields the reduced stochastic process in the strong coupling limit given in 
equations~\ref{equ_reduced_sc_p} and~\ref{equ_reduced_sc_v}, the terms due to effects of order $\epsilon$ are captured in the operator $L_0$.

We may be interested in the limit of $L_0$ in the negligible mass limit ($\kappa^{-1} \to 0$). In this case it is important to remember to make certain assumptions about the relationship of the non-dimensional constants $\kappa$, $\epsilon$, and $\alpha$. In particular, we recall the requirement $\kappa \alpha = 1$, which implies $\epsilon \ll (\kappa \alpha)^{-1}$. This avoids the blow-up of the potential terms in $L_0$ when we wish to take $\kappa^{-1} \to 0$.

\subsection{Strong Coupling with Weak Slip: Higher Order Terms in $\epsilon$}
\label{section_weak_slip}
We now consider the next order correction terms in $\epsilon$.  These first-order terms contain contributions that can be interpreted in the strong coupling regime as the weak leading-order slip-effects between the fluid and structures.  Such effects are known to arise in small-scale systems from hydrophobic effects or a break-down of the continuum hypothesis (non-negligible Knudsen number)~\cite{EricLauga2005,Cunningham1910}.  In the SELM formulation the precise form of the slip arises from the choice of coupling operators $\Lambda$ and $\Gamma$.  To capture weak slip effects, these terms could possibly be used to incorporate leading-order slip effects for analysis or for computational simulations without suffering the rapid dynamics over short time-scales associated with the strong coupling.

For the terms below we have assumed the product $\Gamma \Lambda$ does not depend on $\mathbf{X}$.  This corresponds to the situation where the coupling operators for the particles/microstructure do not overlap significantly (in practical simulations this is often imposed by steric repulsion forces) and the situation where the particle mobility is translation invariant (closely related to the coupling operator product).  If this condition is not the case, similar analysis  would yield additional terms.

For the SELM formulation, it is convenient to express the effective ``slip terms'' $\theta_p$ and $\theta_X$ in Section~\ref{section_summary_strong_coupling} by decomposition into the parts 
\begin{align}
\theta_p = \Theta^p + \Theta^p_{\text{thm}}, \mbox{\hspace{1cm}} 
\theta_X = \Theta^X + \Theta^X_{\text{thm}} .
\end{align}
These were computed using the non-dimensional conventions.  It is most convenient to combine the dimensions to the resulting non-dimensional $\epsilon$-order terms, as dictated by the units of each equation.   That is, we multiply
\begin{eqnarray*}
\Theta^p = \Theta^p_0  \bar \Theta^p , \mbox{\hspace{0.3cm}}
\Theta^X = \Theta^X_0  \bar \Theta^X, \mbox{\hspace{0.3cm}}
\Theta^p_{\text{thm}} = \Theta^p_0  \bar \Theta^p_{\text{thm}},\mbox{\hspace{0.3cm}}
\Theta^X_{\text{thm}} = \Theta^X_0  \bar \Theta^X_{\text{thm}},\mbox{\hspace{0.3cm}} \\
\Theta^p_0 = \left({m_0}/{\tau_k \ell^2}\right),\mbox{\hspace{0.3cm}} 
\Theta^X_0 = \left({\ell}/{\tau_k}\right).\mbox{\hspace{5.6cm}} 
\end{eqnarray*}
The non-dimensional $\epsilon$-order corrections to the drift are given by
\begin{align}
 \bar \Theta^p & =   \epsilon \bigg(
	 r
	+ \frac{1}{2} \nabla_{\mathbf{X}} \sigma : \sigma^T
	+Y + Z - B C^{-1} \Upsilon ^{-1} \gamma(\alpha \kappa) \bigg) \\
\bar \Theta^X &=  
	\epsilon  \left(   - C^{-1} \Upsilon ^{-1} \gamma(\alpha \kappa)  \right). 
\end{align}
In our notation, the constituent terms $\bar \Theta$ are understood to have no dimensions.
The non-dimensional $\epsilon$-order corrections to the noise are given by:
	\begin{eqnarray}
	\left \langle  \bar \Theta^p_\subtxt{thm}(\bar s )
	(\bar \Theta^p_\subtxt{thm})^{\RegAdjoint}(\bar t) \right \rangle 
	& = &\epsilon \left( \sigma \sigma^T + 2 B C^{-1}  \bar\OpFSDissp^{-1} C^{-1} B^T \right) \hspace{0.06cm}\delta(\bar t - \bar s), \\
	\left \langle \bar \Theta^X_\subtxt{thm}(\bar s)(\bar \Theta^X_\subtxt{thm})^{\RegAdjoint}(\bar t) \right \rangle 
	& = & \epsilon \left( 2  C^{-1} \bar \OpFSDissp^{-1} C^{-1} \hspace{0.06cm}\right) \delta(\bar t -\bar  s), \\ 
	\left \langle \bar \Theta^p_\subtxt{thm}(\bar s)(\bar \Theta^X_\subtxt{thm})^{\RegAdjoint}(\bar t) \right \rangle 
	& = & \epsilon \left( 2 B  C^{-1} \bar \OpFSDissp^{-1} C^{-1} \hspace{0.06cm}\right) \delta(\bar t -\bar  s).
	\end{eqnarray}
$q$ was used to construct the above noise terms.
The $r$ and $q$ are given in~\ref{equ_r }.  $Y$, $Z$,  $\sigma$, $B$, and $\gamma$, are given in  equations~\ref{equ_Y},~\ref{equ_omega},~\ref{equ_sigma},~\ref{def_B}, and~\ref{def_gamma}.
A derivation of these equations is given in Section~\ref{sec_derivation_strong_coupling}.

\subsection{Limit of Negligible Excess Mass: Derivation of Reduced Equations}
\label{sec_derivation_small_excess_mass}
To obtain the reduced equations in the limit of negligible excess mass, we 
consider the limit $\kappa ^{-1} \rightarrow 0$, with the limit 
$\epsilon \rightarrow 0$ assumed to be already taken for simplicity. This corresponds to 
the physical regime with $m \ll  m_0 = \rho \ell^3$ with the coupling very strong.

We begin with the $\bar{L}_1$ operator, which describes the dynamics for the limit $\epsilon \to 0$ taken. This operator is given by equation~\ref{bar_L1}. In the limit $\kappa^{-1} \to 0$ we find
\begin{eqnarray}
\label{bar_L1_lim_kappa}
\bar{L}_1 &= & \left( \frac{1}{\delta}  \OpFldDissp \mathbf{p} 
 - \alpha \OpSFCouple \nabla_\mathbf{X} \Eng (\mathbf{X})+ \nabla_{\mathbf{X}} \OpSFCouple : \mathcal{I}
 \right) \cdot \nabla_{\mathbf{p}}
 -\frac{1}{\delta}\OpFldDissp: \nabla_{\mathbf{p}}^2 
 +  {\mathbf{v}_0} \cdot \nabla_{\mathbf{X}}
\end{eqnarray}
Equations~\ref{equ_v0}
simplify with
\begin{align}
\label{equ_v0_lim_kappa}
\mathbf{v}_0 = \OpFSCouple \mathbf{u}, \mbox{\hspace{1cm}}
\nabla_{\mathbf{X}} \OpSFCouple  : \mathcal{I} 
= \text{tr} [\nabla_{\mathbf{X}} \OpSFCouple] = \nabla_{\mb{X}} \cdot \OpSFCouple .
\end{align}
We obtain~\ref{equ_u_bar_m_zero_fld}--\ref{equ_f_thm_m_zero_fld} by adding the units to the non-dimensional variables and writing the equation in its dynamical form.  For clarity we also add the Lagrangian forces and the $\theta$ terms which appear if the slip terms are small but not negligible. 

\subsection{Limit of Rapid Hydrodynamic Relaxation: Derivation of Reduced Equations}
\label{sec_derivation_rapid_fluid_relax}
We now consider the regime where the hydrodynamics relaxes rapidly relative to the time-scale of the microstructure motions.  
We assume for simplicity the regime in which we have already taken $\epsilon \to 0$ and then $\kappa ^{-1} \to 0$.  We consider the non-dimensional group $\delta$ given by ratio of time-scales between hydrodynamic relaxation $\rho \ell^2 / \mu$ and advective transport $\ell / U$.  In particular, $\delta = {(\rho \ell^2 / \mu)}/{(\ell / U)} =  \rho \ell U / \mu $, also defined in equation~\ref{equ_def_delta}, where $U$ is a characteristic flow 
velocity, $\ell$ a characteristic length-scale, and $\mu$ the fluid viscosity.  We consider the limit $\delta \rightarrow 0$.

To handle the incompressibility constraint on the fluid, we now introduce a projection operator approach 
to handle the Lagrange multiplier $\lambda$ in
\begin{eqnarray}
\rho\frac{d\mb{u}}{dt}  & = & \OpFldDissp \mb{u}   
                          - \OpSFCouple \nabla_{\mb{X}}\Phi(\mb{X}) 
                          + k_B T \nabla_{\mb{X}} \cdot \OpSFCouple 
                          + \mb{f}_\subtxt{thm}
				+\lambda \\
\frac{d\mb{X}}{dt}  & = & \OpFSCouple \mb{u}.
\end{eqnarray}
The $\lambda$ acts as a constraint force density that enforces $\nabla \cdot \mb{u} = 0$.  This can be written in terms 
of a projection operator as
\begin{align}
\lambda = -(\mathcal{I} - \proj)
				(\OpFldDissp \mb{u}   
                          - \OpSFCouple \nabla_{\mb{X}}\Phi(\mb{X}) 
                          + k_B T \nabla_{\mb{X}} \cdot \OpSFCouple 
                          + \mb{f}_\subtxt{thm}).
\end{align}
The projection operator is given by $\wp= \mathcal{I} - \nabla \Delta^{-1} \nabla\cdot$.  This gives
\begin{eqnarray}
\rho\frac{d\mb{u}}{dt}  & = & \proj[\OpFldDissp \mb{u}   
                          - \OpSFCouple \nabla_{\mb{X}}\Phi(\mb{X}) 
                          + k_B T \nabla_{\mb{X}} \cdot \OpSFCouple 
                          + \mb{f}_\subtxt{thm} ]\\
\frac{d\mb{X}}{dt}  & = & \OpFSCouple \mb{u}.
\end{eqnarray}
We assume that $\proj \OpFldDissp = \OpFldDissp \proj$ and make use of the properties $\proj^2 = \proj$ and $\proj = \proj^T$.
We can express this in non-dimensionalized form using $\delta$ as defined in equation~\ref{equ_def_delta} as
\begin{eqnarray}
\frac{d\mb{u}}{dt}  & = & \proj \bigg[\frac{1}{\delta}\OpFldDissp \mb{u}   
                          - \OpSFCouple \nabla_{\mb{X}}\Phi(\mb{X}) 
                          + \nabla_{\mb{X}} \cdot \OpSFCouple 
                          + \sqrt{\frac{1}{\delta}} \mb{f}_\subtxt{thm} \bigg]\\
\frac{d\mb{X}}{dt}  & = & \OpFSCouple \mb{u}.
\end{eqnarray}
The infinitesimal generator is given by 
\begin{eqnarray}
\mathcal{A} &=&
				\frac{1}{\delta} [\proj \OpFldDissp \mathbf{u} 
				+ \delta \proj (- \OpSFCouple \nabla_{\mb{X}}\Phi(\mb{X}) 
                          + \nabla_{\mb{X}} \cdot \OpSFCouple ) ] \cdot \nabla_{\mathbf{{u}}} - \frac{1}{\delta}( \proj \OpFldDissp): \nabla^2_{\mathbf{{u}}} 
				+ [  \OpFSCouple \mb{u} ] \cdot \nabla_{\mathbf{X}}.
\end{eqnarray}
To apply our perturbation analysis introduced in Section~\ref{sec_reduction}, we split the operator as
\begin{eqnarray}
\mathcal{A} = L_{slow} + L_{fast}, \mbox{\hspace{0.3cm}} 
L_{slow} = \bar{L}_1 + L_1, \mbox{\hspace{0.3cm}} 
L_{fast} = L_2 + \tilde{L}_2, \\
\bar{L}_1  = 0, \mbox{\hspace{0.3cm}}  
L_1 = (\OpFSCouple \mb{u}) \cdot \nabla_{\mathbf{X}}, \mbox{\hspace{3cm}} \\  
L_2 = (\proj \OpFldDissp \mb{u})\cdot \nabla_{\mathbf{u}} 
				-(\proj \OpFldDissp) : \nabla^2_\mathbf{u}, \mbox{\hspace{0.3cm}}  
\tilde{L}_2 =  \proj (- \OpSFCouple \nabla_{\mb{X}}\Phi(\mb{X}) 
                          + \nabla_{\mb{X}} \cdot \OpSFCouple ) \cdot \nabla_\mathbf{u}.
\end{eqnarray}
We remark that each of the coefficients of $L_2$ and $\tilde{L}_2$ are  in the range of $\proj$.  Using the linearity of 
$\proj$ and $\OpFldDissp$ we may interpret the derivatives $\nabla_\mathbf{u}$ as $\nabla_\mathbf{\proj u}$ and complete the reduction with the fast 
variable in the space $u \in \mathcal{S}$.  The $\mathcal{S}$ denotes our space of solenoidal vector fields.  Given the 
specific form of $L_2$, the inverse operator can be expressed as
\begin{align}
L_2^{-1} \mathbf{u} = \OpFldDissp^{-1} \mathbf{u} \text{, \hspace{0.5cm} for } u\in\mathcal{S}.
\end{align}
This allows us to carry out readily the inverse 
\begin{align}
L_1 L_2^{-1} L_1& 
= L_1L_2^{-1} [(\OpFSCouple \mb{u}) \cdot \nabla_{\mathbf{X}}] 
= L_1 [(\OpFSCouple \OpFldDissp^{-1} \mb{u}) \cdot \nabla_{\mathbf{X}}]\\
&= (\OpFSCouple \mb{u}) \cdot \nabla_{\mathbf{X}} [(\OpFSCouple  \OpFldDissp^{-1} \mb{u}) \cdot \nabla_{\mathbf{X}}] 
\text{, \hspace{0.5cm} for }u\in\mathcal{S}.
\end{align}
We still need to evaluate $\int_{u\in\mathcal{S}} \psi(\mathbf{u}) L_1 L_2^{-1} L_1$.  An important feature is that the covariance structure of $\psi$ is the identity in the space $\mathcal{S}$ because the operator coefficients on the first and second order terms in $L_2$ are identical.  It is useful to rewrite the inverse as
\begin{align}
L_1 L_2^{-1} L_1
&= (\OpFSCouple \mb{u}) \cdot \nabla_{\mathbf{X}} [(\OpFSCouple  \OpFldDissp^{-1} \mb{u}) \cdot \nabla_{\mathbf{X}}]\\
&= \nabla_{\mathbf{X}} \cdot \{ (\OpFSCouple \mb{u}) [(\OpFSCouple  \OpFldDissp^{-1} \mb{u}) \cdot \nabla_{\mathbf{X}}]\}
-   (\nabla_{\mathbf{X}} \cdot (\OpFSCouple \mb{u})) [(\OpFSCouple  \OpFldDissp^{-1} \mb{u}) \cdot \nabla_{\mathbf{X}}].
\end{align}
The averaging with respect to $\psi(\mathbf{u})$ can be computed readily in this form by passing the integral onto $\mathbf{u}$ inside each term. 
From the covariance structure of $\psi$ we have the useful identities $\int u_i u_j \psi(\mathbf{u}) d\mathbf{u} = \delta_{ij}$ 
for $u \in \mathcal{S}$.  By using these identities and that $\OpFSCouple = \OpSFCouple^T$, we have 
\begin{align}
& \int_{u\in\mathcal{S}} \psi(\mathbf{u}) L_1 L_2^{-1} L_1
=  \nabla_{\mathbf{X}} \cdot  [(\OpFSCouple  \OpFldDissp^{-1} \proj \OpSFCouple) \cdot \nabla_{\mathbf{X}}]
-   (\OpFSCouple  \OpFldDissp^{-1} \proj \nabla_\mathbf{X} \cdot \OpSFCouple) \cdot \nabla_{\mathbf{X}} \\
&= [\nabla_{\mathbf{X}} \cdot  (\OpFSCouple  \OpFldDissp^{-1} \proj \OpSFCouple)] \cdot \nabla_{\mathbf{X}}
+  (\OpFSCouple  \OpFldDissp^{-1} \proj \OpSFCouple) : \nabla^2_{\mathbf{X}}
-   (\OpFSCouple  \OpFldDissp^{-1} \proj \nabla_\mathbf{X} \cdot \OpSFCouple) \cdot \nabla_{\mathbf{X}}.
\end{align}
We can evaluate the remaining integral term in $L_0$ by 
\begin{align}
&\int_{u\in\mathcal{S}} \psi(\mathbf{u}) \tilde{L}_2 L_2^{-1} L_1 \\
&= \int_{u\in\mathcal{S}} \psi(\mathbf{u}) \proj(- \OpSFCouple \nabla_{\mb{X}}\Phi(\mb{X}) 
                          + \nabla_{\mb{X}} \cdot \OpSFCouple ) \cdot \nabla_\mathbf{u}
			 [(\OpFSCouple  \OpFldDissp^{-1} \mb{u}) \cdot \nabla_{\mathbf{X}}] \\
&= \int_{u\in\mathcal{S}} \psi(\mathbf{u})
			 [(\OpFSCouple  \OpFldDissp^{-1} \proj (- \OpSFCouple \nabla_{\mb{X}}\Phi(\mb{X}) 
                          + \nabla_{\mb{X}} \cdot \OpSFCouple )) \cdot \nabla_{\mathbf{X}}] \\
&= (\OpFSCouple  \OpFldDissp^{-1} \proj (- \OpSFCouple \nabla_{\mb{X}}\Phi(\mb{X}) 
                          + \nabla_{\mb{X}} \cdot \OpSFCouple )) \cdot \nabla_{\mathbf{X}}.
\end{align}
For the effective infinitesimal operator $\bar{L} = \bar{L}_1 + \delta \cdot L_0$, this gives
\begin{align}
\bar{L} &= -\delta\cdot\left( \int_{u\in\mathcal{R}(\proj)} \psi(\mathbf{u}) \tilde{L}_2 L_2^{-1} L_1 + \int_{u\in\mathcal{R}(\proj)} \psi(\mathbf{u}) L_1 L_2^{-1} L_1  \right) \\
& =  - \delta\cdot \bigg\{ (
 \OpFSCouple  \OpFldDissp^{-1} \proj [  
				- \OpSFCouple \nabla_{\mb{X}}\Phi(\mb{X}) 
                          + \nabla_{\mb{X}} \cdot \OpSFCouple ]) \cdot \nabla_{\mathbf{X}} \\
&+ [\nabla_{\mathbf{X}} \cdot  (\OpFSCouple  \OpFldDissp^{-1} \proj \OpSFCouple)] \cdot \nabla_{\mathbf{X}}
+  (\OpFSCouple  \OpFldDissp^{-1}\proj \OpSFCouple) : \nabla^2_{\mathbf{X}}
-   (\OpFSCouple  \OpFldDissp^{-1} \proj \nabla_\mathbf{X} \cdot \OpSFCouple) \cdot \nabla_{\mathbf{X}} \bigg \} \nonumber \\
&= \delta\cdot \bigg\{ \bigg[
 \OpFSCouple (- \OpFldDissp)^{-1} \proj [  
				- \OpSFCouple \nabla_{\mb{X}}\Phi(\mb{X}) 
				 ]
		+ [\nabla_{\mathbf{X}} \cdot  (\OpFSCouple (- \OpFldDissp)^{-1}\proj \OpSFCouple)] \bigg] \cdot \nabla_{\mathbf{X}} \\
		&+  (\OpFSCouple (- \OpFldDissp)^{-1} \proj \OpSFCouple) : \nabla^2_{\mathbf{X}} \bigg\}. \nonumber
\end{align}

By letting $\tilde{H}_{\subtxt{SELM}} =  -\OpFSCouple \OpFldDissp^{-1} \proj \OpSFCouple $, we can express this more compactly as 
\begin{align}
\bar{L}
&= \delta\cdot \bigg\{ \bigg[
\tilde{H}_{\subtxt{SELM}}(-\nabla_{\mb{X}}\Phi(\mb{X}) ) 
		+\nabla_\mathbf{X} \cdot \tilde{H}_{\subtxt{SELM}}\bigg]
 \cdot \nabla_{\mathbf{X}}
		+  \tilde{H}_{\subtxt{SELM}} : \nabla^2_{\mathbf{X}} \bigg \}.
\end{align}

By converting the expressions to have physical units, the reduced stochastic processes in the limit of fast hydrodynamic relaxation becomes
\begin{eqnarray}
\frac{d\mathbf{X}}{dt} =
		 H_{\subtxt{SELM}} [-\nabla_{\mb{X}}\Phi(\mb{X})]
		+ (\nabla_\mathbf{X} \cdot H_{\subtxt{SELM}}) k_B T 
		+\mathbf{h}_{\subtxt{thm}}  \mbox{\hspace{1.3cm}}\\
H_{\subtxt{SELM}} =  \OpFSCouple (-  \OpFldDissp)^{-1} \proj \OpSFCouple,  \mbox{\hspace{0.3cm}}
\langle \mathbf{h}_{\subtxt{thm}}(s), \mathbf{h}^T_{\subtxt{thm}}(t) \rangle = 2 k_B T H_{\subtxt{SELM}} \delta(t-s).
\end{eqnarray}
The properties of $\proj$ allow us to write $H_{\subtxt{SELM}} =  \OpFSCouple \proj^T (-  \OpFldDissp)^{-1} \proj \OpSFCouple $.
This provides a convenient way to factor the hydrodynamic coupling tensor and generate stochastic driving fields.  In particular, 
$H_{\subtxt{SELM}} = Q^T Q$ with $Q = \OpFSCouple \proj^T \sqrt{(-  \OpFldDissp)^{-1} }$.  This can be used to obtain efficient computational methods as in~\cite{Atzberger2014_sFEM}.

\section{Conclusions}
We have shown how to systematically reduce 
descriptions of fluid-structure interactions subjected to thermal fluctuations.  
The presented analysis and reduced equations provide connections 
between the physical regimes often encountered in widely used fluid-structure models.  
We have shown in the strong coupling limit 
that important drift terms arise in the stochastic dynamics that 
were previously neglected in more phenomenological approaches.
To gain insights into the choice of approximate fluid-structure 
interaction operators, we have also shown how effective equations 
can be derived in the overdamped limit to obtain an effective 
hydrodynamic coupling tensor for microstructure responses.
The presented analysis and reduced equations remove extraneous 
degrees of freedom and potential sources of numerical stiffness.
The reduction methods provide a promising approach for the 
further development of descriptions and computational methods 
for fluid-structure interactions when subject to thermal 
fluctuations.

\section{Acknowledgements}
The author P.J.A. and G.T. acknowledges support from research 
grant NSF CAREER DMS - 0956210 and DOE CM4.  G.T also
acknowledges undergraduate support from NSF REU DMS-0852065
and  from the UCSB CCS program, and graduate support from
 the Department of Defense (DoD) through the National Defense Science \& Engineering Graduate Fellowship (NDSEG) Program 
and the Stanford Graduate Fellowship.

\bibliography{paper_database}

\appendix

\section{Inverting $L_2$ : General Method from Sturm-Liouville Theory}
\label{sec_sl}

In the stochastic reduction procedure of Section~\ref{sec_reduction} the operator $L_2$ needs to be inverted.  
In general, this presents one of the most significant 
challenges in determining the form of the reduced equations.  Interestingly in the regimes we consider for the SELM stochastic 
dynamics, the $L_2$ operators can be related to Sturm-Liouville problems~\cite{Strauss2008} and the inversions can be 
represented readily over a finite sum over elements of an orthonormal basis of known functions. 
Here we develop this general theory, which can be used as an effective method to invert the $L_2$ operators that arise in  practice for the SELM stochastic dynamics.

The $L_2$ is an infinitesimal generator of a stochastic process and second order differential operator.
We refer to it as a \text{Sturm-Liouville operator} (for short a SL-operator) if it has the specific form
\begin{eqnarray}
L_2 = -\frac{\partial}{\partial z} \left\lbrack p(z) \frac{\partial}{\partial z} \right\rbrack + q(z)
\end{eqnarray}
where $p(z) >0$ is continuously differentiable and $q(z) \geq 0$ is continuous.
In practice, not all of the infinitesimal generators we 
encounter will be SL-operators directly but with a change of variable can be related to an 
operator of this form.  
The inverses we encounter are 
typically of the form $w = -L_2^{-1} L_1 u $ which amounts to finding a solution to the 
problem 
\begin{eqnarray}
L_2 w = f(z).
\end{eqnarray}
As we shall discuss the functions $f(z)$ that arise in practice often have special properties
that we can utilize.  The Sturm-Liouville property has the important consequence that an
orthonormal basis of eigenfunctions $\phi_n$ can be constructed for this operator by solving
\begin{eqnarray}
\label{equ_un_eig}
L_2 \phi_n = \lambda_n \mu(z) \phi_n.
\end{eqnarray}
We also require that the solution satisfy the decay condition $|\phi_n| \to 0$ as $|z| \to \infty$.
The $\mu$ is a fixed continuous function with $\mu(z) > 0$ which defines a weighted inner product for which $L_2$ is self-adjoint 
\begin{eqnarray}
\langle g(z),h(z) \rangle_\mu & = & \int g(z) \overline{h(z)} \mu(z) dz.
\end{eqnarray}
This defines a Hilbert space ${L}^2(\mu)$.
Since $\phi_n$ are eigenfunctions of $L_2$ in the sense of equation~\ref{equ_un_eig}, it is useful to express the inverse problem as
\begin{eqnarray}
\label{equ_inverse_problem_sl}
L_2 w = \tilde{f}(z) \mu(z)
\end{eqnarray}
where $\tilde{f}(z) = {f(z)}/{\mu(z)}$.  Assuming $\tilde{f}(z),w(z) \in L^2(\mu)$ and satisfy the decay condition, 
the orthonormal basis can then be used to represent both $w$ and 
$\tilde{f}$ as 
\begin{eqnarray}
\tilde{f}(z) = \sum_k \tilde{f}_k \phi_k(z), \mbox{\hspace{1cm}} 
w(z) = \sum_k w_k \phi_k(z) \\
\tilde{f}_k  = \langle \tilde{f}(z), \phi_k(z) \rangle_\mu, \mbox{\hspace{1cm}} 
w_k  = \langle w(z), \phi_k(z) \rangle_\mu.
\end{eqnarray}
Plugging this into equation~\ref{equ_inverse_problem_sl} gives
\begin{eqnarray}
\langle L_2 w, \phi_k(z) \rangle = w_k \lambda_k = \tilde{f}_k. 
\end{eqnarray}
The $\langle \cdot, \cdot \rangle$ without the subscript denotes the usual 
unweighted $L^2$-inner-product.  For the inverse $w = L_2^{-1} f$, this 
provides the following representation
\begin{eqnarray}
w(z) = \sum_k w_k \phi_k(z), \mbox{\hspace{1cm}}  w_k  = \tilde{f}_k/\lambda_k. 
\end{eqnarray}

\section{Table of constants}

\begin{small}
\begin{tabularx}{\textwidth}{ |c|X| }
  \hline
 Value & Description  \\
  \hline
  $\OpFSDissp_0$  & Fluid-structure momentum coupling constant.   \\
    
 $\rho$ & Density of the fluid.  \\
  
 $\ell$ & Characteristic structure length-scale.  \\

 $m$ & Excess mass of a structure.  \\

 $m_0 = \rho \ell^3$ & Mass of displaced fluid within volume $\ell^3$.  \\
  
 $k_B T$ & Thermal energy.  \\
  
 $\tau_v = \OpFSDissp_0 / m$  & Microstructure velocity relaxation time-scale.  \\
  
 $\tau_k = \sqrt{m_0 \ell^2/k_B{T}} $ &   Time-scale for diffusion over distance $\ell$. \\
    
$\kappa = \rho \ell^3 / m = m_0 / m$ & Ratio of fluid mass to excess structure mass. \\
  
$\epsilon = \tau_v/\tau_k$ & Ratio of inertial time-scale to diffusive time-scale. \\

$\alpha = \Phi_0 / k_B T$ & Characteristic potential energy relative to thermal energy.\\
  
  \hline
\end{tabularx}
\newline
\end{small}

\end{document}